%% file: main.tex
\documentclass{article}

\input{preamble}

\title{Visored: A Controlled-Natural-Language Prover for LLM-Generated Mathematics}

\author{%
 Xiyu Zhai\thanks{University of Washington. Correspondence to Xiyu Zhai: \texttt{xiyuzhai@cs.washington.edu}} \quad
 Xinyi Chen\footnotemark[1] \quad
 Yiping Wang\footnotemark[1] \\
 Runlong Zhou\footnotemark[1] \quad
 Liao Zhang\thanks{University of Innsbruck} \quad
 Simon S. Du\footnotemark[1]
}

\date{}  

\begin{document}
\maketitle

\input{abs_intro}

\input{sections/architecture}

\input{sections/worked-example}

\input{sections/intended-usage}

\input{sections/experiments}

\input{sections/limitations}

\bibliographystyle{plain}
\bibliography{references/cnl,references/llm-autoformalization,references/llm-general,references/ai-math,references/foundations,references/tactic-layers,references/proof-checkers,references/pvs,references/benchmarks,references/coding-agents}

\clearpage

\appendix

\input{appendix/related-work}

\input{appendix/syntax}

\input{appendix/semantics}

\input{appendix/kernel}

\input{appendix/solver}

\input{appendix/transpilation}

\input{appendix/skill}

\input{appendix/examples}

\end{document}

%% file: preamble.tex
\usepackage[english]{babel}

\usepackage[letterpaper,top=2cm,bottom=2cm,left=3cm,right=3cm,marginparwidth=1.75cm]{geometry}

\usepackage{amsmath}
\usepackage{amssymb}
\usepackage{amsthm}
\usepackage{mathtools}

\usepackage{graphicx}
\usepackage{booktabs}
\usepackage{listings}
\usepackage{xcolor}
\usepackage{tikz}
\usetikzlibrary{calc, arrows.meta, positioning, shapes, arrows}

\usepackage{enumitem}
\setlist{nosep,leftmargin=*}

\usepackage[colorlinks=true, allcolors=blue]{hyperref}

\usepackage{iftex}
\ifxetex
  \usepackage{xeCJK}
\fi

\newtheorem{example}{Example}

\lstdefinelanguage{Visored}{
  morecomment=[l]{\%},
  basicstyle=\ttfamily\footnotesize,
  keywordstyle=\color{blue!70!black}\bfseries,
  morekeywords={begin,end,example,Let,Assume,The,goal,is,to,prove,Then,We,have},
  showstringspaces=false,
  breaklines=true,
  columns=fullflexible,
}

\lstdefinelanguage{Lean4}{
  morecomment=[l]{--},
  morecomment=[s]{/-}{-/},
  basicstyle=\ttfamily\footnotesize,
  keywordstyle=\color{purple}\bfseries,
  morekeywords={theorem,lemma,def,by,exact,intro,apply,have,let,fun,if,then,else,match,with,return,do,where,instance,namespace,end,example,import,subst,nlinarith,norm_num,linarith,ring,rw,refine,obtain,show,sorry},
  showstringspaces=false,
  breaklines=true,
  columns=fullflexible,
  extendedchars=true,
  inputencoding=utf8,
  literate={⁻¹}{{${}^{-1}$}}2
           {⁻²}{{${}^{-2}$}}2
           {⁻³}{{${}^{-3}$}}2
           {ℝ}{{$\mathbb{R}$}}1
           {ℕ}{{$\mathbb{N}$}}1
           {ℤ}{{$\mathbb{Z}$}}1
           {ℚ}{{$\mathbb{Q}$}}1
           {ℂ}{{$\mathbb{C}$}}1
           {≤}{{$\le$}}1
           {≥}{{$\ge$}}1
           {≠}{{$\ne$}}1
           {≡}{{$\equiv$}}1
           {→}{{$\to$}}1
           {←}{{$\leftarrow$}}1
           {↔}{{$\leftrightarrow$}}1
           {↦}{{$\mapsto$}}1
           {⇒}{{$\Rightarrow$}}1
           {∀}{{$\forall$}}1
           {∃}{{$\exists$}}1
           {∈}{{$\in$}}1
           {∉}{{$\notin$}}1
           {∧}{{$\wedge$}}1
           {∨}{{$\vee$}}1
           {¬}{{$\neg$}}1
           {⟨}{{$\langle$}}1
           {⟩}{{$\rangle$}}1
           {·}{{$\cdot$}}1
           {⁰}{{${}^{0}$}}1
           {¹}{{${}^{1}$}}1
           {²}{{${}^{2}$}}1
           {³}{{${}^{3}$}}1
           {⁴}{{${}^{4}$}}1
           {⁵}{{${}^{5}$}}1
           {₀}{{${}_{0}$}}1
           {₁}{{${}_{1}$}}1
           {₂}{{${}_{2}$}}1
           {₃}{{${}_{3}$}}1
           {₄}{{${}_{4}$}}1
           {₅}{{${}_{5}$}}1
           {⁻}{{${}^{-}$}}1
           {∑}{{$\sum$}}1
           {∏}{{$\prod$}}1
           {∘}{{$\circ$}}1
           {√}{{$\surd$}}1
           {∣}{{$\mid$}}1
           {∤}{{$\nmid$}}1
           {∪}{{$\cup$}}1
           {⊆}{{$\subseteq$}}1
           {⌊}{{$\lfloor$}}1
           {⌋}{{$\rfloor$}}1
           {−}{{$-$}}1
           {–}{{--}}1
           {—}{{---}}1
           {‐}{{-}}1
           {•}{{$\bullet$}}1
           {…}{{$\ldots$}}1
           {─}{{-}}1
           {α}{{$\alpha$}}1
           {β}{{$\beta$}}1
           {γ}{{$\gamma$}}1
           {δ}{{$\delta$}}1
           {ε}{{$\varepsilon$}}1
           {λ}{{$\lambda$}}1
           {π}{{$\pi$}}1
           {σ}{{$\sigma$}}1
           {τ}{{$\tau$}}1
           {φ}{{$\varphi$}}1
           {ω}{{$\omega$}}1,
}

\providecommand{\yiping}[1]{}

%% file: abs_intro.tex


\begin{abstract}
We present a dependent-type-based prover designed around the way LLMs (and humans) tend to write mathematics, complementing existing systems such as Lean and Rocq. Its core design choices are a surface that imitates mathematical natural language and a rule-driven automation layer that closes the routine steps a textbook would omit, so that an accepted proof can be re-emitted as a checked Lean file. Early experiments suggest that, even without any prover-specific training data, LLMs can learn to use it effectively on the miniF2F benchmark. Lean output excerpts: \url{ https://github.com/xiyuzhai-husky-lang/visored/}
\end{abstract}

\input{intro_trimmed}


\section{Related Work}
\label{sec:related-work}

Visored draws together several earlier schools of prover design. It shares the controlled-natural-language surface of Mizar and Naproche~\cite{mizar,naproche2021}, the typecheck-time well-definedness of PVS~\cite{pvs1992,pvs-subtypes}, the dependently-typed substrate of Lean and Rocq, and the cost-budgeted rule-driven solving of ProofGrader~\cite{proofgrader}, and it adopts the LLM-centric framing of the recent autoformalization literature~\cite{wu2022autoformalization,jiang2023dsp,xin2024deepseekprover}. We do not claim any individual ingredient is novel; what is new is their integration into a single LLM-facing pipeline whose checkable intermediate representation makes failures localized rather than a single accept/reject after a Lean compilation. Appendix~\ref{app:related-work} gives the full discussion: the broader progress of AI for mathematics that motivates the autoformalization bottleneck, and a detailed comparison with each prior school, namely direct LLM autoformalization and whole-proof Lean models, CNL provers, natural-language tactic layers, rule-based proof checkers, and typecheck-time well-definedness.

%% file: intro_trimmed.tex
\section{Introduction}

Large language models (LLMs) have become increasingly capable mathematical reasoners, both on standard benchmarks and, more recently, at the level of competition and research-level mathematics. They nevertheless still suffer from hallucination~\cite{ji2023hallucination,huang2023hallucination}, so an AI-produced proof is only as trustworthy as the process that checks it, and the most striking recent results still relied on substantial human verification before they could be reported. As more mathematics is produced with AI assistance, the binding constraint shifts from \emph{generating} ideas to \emph{verifying} and \emph{organizing} them at scale. Autoformalization, the task of mapping informal mathematical writing into machine-checkable form, is one concrete instance of that bottleneck: if AI-generated proofs are to be absorbed by the mathematical community rather than accumulating as unverified candidates, the route from informal prose to a kernel-checked proof has to become much more reliable than it currently is. This paper presents one attempt in that direction.

One line of work removes the human from the loop with a small, auditable kernel that returns a definite yes/no on a candidate proof, independent of the model that produced it (the de Bruijn criterion~\cite{debruijn-criterion,paulson2022lcf}). Whole-proof theorem provers now drive miniF2F~\cite{minif2f} pass rates close to saturation (Appendix~\ref{app:related-work}), but they take an \emph{already formalized} statement as input. Turning informal prose into that statement and its proof, the autoformalization step itself, lags well behind~\cite{weng2025autoformalizationsurvey} and is where most of the non-local surface-to-semantics decisions are made.

The separation is not just a user-experience (UX) issue. Informal and formal mathematics are not identical artifacts: informal prose leans on shared background, controlled abuse of notation, implicit side conditions (``for $x$ sufficiently large'', ``WLOG $x \neq 0$''), and implicit number-system and coercion choices ($\mathbb{N}$ vs.\ $\mathbb{Z}$ vs.\ $\mathbb{R}$) that readers fill in automatically. A direct translation must therefore commit to many decisions with no surface trace in the prose: the right library lemma, subtype, edge-case convention (how a library defines $1/0$ or $0^0$), tactic, and integer/rational cast. Each is non-local, and getting one wrong rarely surfaces as a compiler error (those are quick to fix) but as a silent semantic drift: a statement that compiles yet no longer says what the prose claimed, or a goal that looks plausible but is unprovable for a reason several rewriting steps upstream.

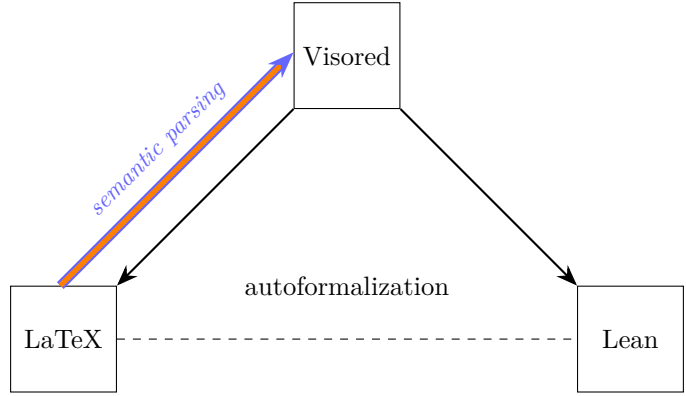
\begin{figure}[t]
\centering
\begin{tikzpicture}[
    >=Stealth,
    node distance=3.75cm,
    every node/.style={draw, text centered, minimum size=1.4cm},
    scale=1.5
]
\node (visored) at (0, 2.5) {Visored};
\node (latex) at (-2.5, 0) {LaTeX};
\node (lean) at (2.5, 0) {Lean};

\draw[thick, -{Stealth[length=8pt]}] (visored) -- (latex);
\draw[thick, -{Stealth[length=8pt]}] (visored) -- (lean);
\draw[thick, color=blue!60, -{Stealth[length=10pt]}, double=orange, double distance=2pt]
    ($(latex.north east)+(-0.5,0)$) --
    node[midway, above=-10pt, draw=none, sloped, font=\small\itshape\color{purple}] {semantic parsing}
    ($(visored.south west)+(0,0.5)$);
\draw[dashed] (latex) -- node[midway, above, draw=none] {autoformalization} (lean);
\end{tikzpicture}
\caption{The Visored triangle. Visored carries the full syntactic, semantic, and logical content of a proof; the maps Visored\,$\to$\,LaTeX and Visored\,$\to$\,Lean are straightforward projections. The task we want to solve, the dashed LaTeX\,$\to$\,Lean edge, is \emph{autoformalization}. The proposed route is the composition of the upward \emph{semantic parsing} arrow (inverting the LaTeX projection) with the deterministic Visored\,$\to$\,Lean projection.}
\label{fig:triangle}
\end{figure}

Our project \textsc{Visored} is a research prototype aimed at making autoformalization as easy and reliable as possible.
Figure~\ref{fig:triangle} states the design thesis. Visored is a semantically rich intermediate representation (IR) in which every decision that has no surface trace in the LaTeX, namely subtype, library lemma, edge-case convention, and cast, is made explicit. The downward maps Visored\,$\to$\,LaTeX (keep the surface, drop the typing) and Visored\,$\to$\,Lean (emit each construct as the corresponding Lean term) are straightforward projections, so the hard part of autoformalization is the upward \emph{semantic parsing} arrow from LaTeX into Visored. That ascent is an \emph{ill-posed inverse problem}: the projection to LaTeX is lossy, many Visored expressions share the same surface, and choosing among them requires a prior over what mathematicians (and LLMs) tend to mean. Supplying that prior is the role of the LLM in the loop; Visored handles everything mechanical below the lift and surfaces failures as localized diagnostics rather than a single accept/reject at the end of a Lean compilation.

Two observations shape the design:

\paragraph{1. LLM math data is overwhelmingly informal.} Most modern mathematics has not been formalized: Mathlib, the Coq stdlib, and Isabelle developments together cover only a small fraction of the field. Frontier LLMs accordingly see far more informal mathematics in training (textbooks, papers, lecture notes) than formal. A natural-language surface stays close to the side the model is already fluent on.

\paragraph{2. Bridging natural and formal reasoning is more tractable than full automated theorem proving.} Traditional automated theorem proving (ATP), namely SAT/SMT solvers and resolution provers, targets fully automated proof discovery, which is NP-hard or undecidable. Visored targets a different problem: with the LLM supplying the proof outline in controlled natural language (CNL), the solver only has to close the small obvious steps that mathematicians routinely omit, a much more tractable task that is well within reach of modern agentic engineering, as demonstrated by coding agents such as Cursor~\cite{cursor}, Claude Code~\cite{claudecode}, and OpenAI Codex~\cite{openaicodex}.

Concretely, Visored takes as input a \textbf{controlled subset of mathematical English embedded in LaTeX} (Appendix~\ref{sec:syntax}). The accepted subset, both the sentence templates (``Let \dots'', ``Assume \dots'', ``Then \dots'') and the LaTeX commands they wrap, is defined by external spec files rather than hard-coded into the parser, so widening or specializing the language is a matter of editing configuration. The input is elaborated into a dependently-typed IR (Appendix~\ref{sec:semantics}), and each ``Then \dots'' obligation is decided by a cost-bounded rule-driven solver (Appendix~\ref{sec:solver}). Following PVS, the elaborator enforces well-definedness at typecheck time: $1/x$ is accepted only when $x \neq 0$ is derivable, $\sqrt{x}$ requires $x \ge 0$, and $\log x$ requires $x > 0$. Accepted proofs can optionally be re-expressed as Lean 4 for kernel-level re-verification, but Visored does not route its own verdict through Lean. Section~\ref{sec:intended-usage} lays out the resulting usage modes: inference-time verifier, autoformalizer loop, dense reward signal, and data generator.

Two challenges sit underneath this design. On the \textbf{ATP side}, Visored needs a solver strong enough to reliably close the small obvious steps that mathematicians omit across arithmetic, algebra, ordering, set theory, and named-lemma applications; without it the workflow stalls in retry loops on steps a human reader would skim past. On the \textbf{ITP side}, producing the dependent-type-checked proofs that interactive theorem provers (ITP) such as Lean and Rocq require is hard, since type systems, coercions, library naming, and partial-function side conditions must all be resolved before the kernel accepts the proof; a recent Claude-Code case study found Lean proof emission to be ``the most challenging aspect'' of building an SMT solver from scratch~\cite{llm2smt2026}, and Naproche has listed ForTheL\,$\to$\,Lean as a direction since 2020 without yet making it a production path~\cite{naproche-sad2020}.

Visored today is an early prototype around the miniF2F dataset, covering middle-school-level set theory, algebra, and inequalities; we discuss the limits openly in Section~\ref{sec:limitations}. It draws together several earlier schools of prover design, reviewed in Appendix~\ref{app:related-work}. What is genuinely new lives in the integration details, namely implicit hypothesis arguments per user expression discharged by the solver, a spec-driven extensible LaTeX surface, a localized per-stage diagnostic channel, and the data structures connecting these into one pipeline, rather than in any individual layer. What we test is whether the combination amounts to more than the sum of its parts when the user is an LLM.

%% file: sections/architecture.tex
\section{Architecture}
\label{sec:architecture}

\input{sections/arch_pipeline_fig}

Visored is itself a prover. Given a LaTeX document (proof content lives inside \texttt{example} environments) plus a small set of config and spec files, it returns its own verdict --- accepted, or a structured diagnostic pointing at the specific source location where elaboration or the solver failed. The Lean emitter is an optional downstream stage that re-expresses an accepted proof as a Lean file for users who need external verification or interoperability with the Lean ecosystem; Visored's correctness does not pass through Lean.

The pipeline (Figure~\ref{fig:pipeline}) is a sequence of stages where each stage either lowers the previous stage's output into a more constrained form or refuses with a diagnostic. We describe each stage by what data it produces and what an LLM gains from being able to inspect that data.

\paragraph{LaTeX source.} The input is LaTeX. Visored accepts a controlled subset of mathematical English written inside LaTeX --- not arbitrary prose --- but the subset is defined by an external, easily editable spec rather than baked into the parser. The set of recognized sentence templates, LaTeX commands, and LaTeX environments lives in \texttt{.lpcsv} configuration files, so extending the surface language is a matter of adding entries (Appendix~\ref{sec:syntax}), not modifying compiler code. In practice the vocabulary covers what an LLM is likely to emit when asked to write a proof in LaTeX, but inputs outside that vocabulary are rejected with a localized diagnostic, not silently coerced.

\paragraph{Syntax tree.} The Syntax stage uses two parsing strategies depending on which mode the source is in. \textbf{Math mode} (the contents of \verb|$...$| and \verb|\[...\]|) is parsed with a precedence-aware stack-based parser similar to a C expression parser: incomplete sub-expressions are pushed onto a stack with their precedences, and the stack is reduced when a closing delimiter or lower-precedence operator arrives. \textbf{Text mode} (CNL prose between math --- ``Let \dots'', ``Assume \dots'', ``Then \dots'') is parsed by trie-like matching of the word/punctuation token stream against the sentence templates declared in spec files (\verb|.lpcsv|): shared-prefix templates are resolved by descending through the alternatives until exactly one continues to match. Both strategies map each LaTeX command (\verb|\sin|, \verb|\frac|, \verb|\sum|, \verb|\mathbb|, \dots) and each LaTeX environment (\verb|example|, \verb|itemize|, \dots) onto a typed syntactic node. The resulting tree is untyped in the type-theoretic sense --- it records what was written, not yet what it means. A LaTeX command the spec does not declare (the user writes \verb|\foobar|) is rejected at this stage with a structured \texttt{UnknownCommand} error pointing at the offending token, which an LLM can patch by rephrasing or by adding the command to the spec.

\paragraph{Typed syntax tree (sem AST).} Names get resolved against the current scope (local definitions first, then global symbols loaded from \texttt{.lpcsv} spec files), and each subexpression is assigned a type and a kind. The kind layer is small and dependent-flavored --- enough for analysis, set theory, basic algebra, and number theory --- and forms the substrate for the typed pattern matching used by later stages.

\paragraph{MIR (mid-level IR).} The sem AST (a typed abstract syntax tree) is lowered to Visored MIR, the substrate the solver and elaborator operate on. A separate stage further translates MIR plus the recorded derivation steps to UVL MIR, a target-neutral IR from which a Lean formatter emits proof text. The split means improvements to the solver do not change emission, and adding a new prover target is a new UVL formatter rather than a rewrite of elaboration.

\paragraph{Elaboration and solver.} Each statement (``Let \dots'', ``Assume \dots'', ``Then \dots'') updates a prover state of variables, hypotheses, and derivations. ``Then \dots'' clauses are the gaps the solver must discharge. Elaboration also enforces well-definedness: writing $1/x$ demands $x \neq 0$, $\sqrt{x}$ demands $x \ge 0$, $\log x$ demands $x > 0$, with the demands flowing asymmetrically through boolean connectives so that a guard like $x \neq 0 \land 1/x > 0$ is accepted. These demands are routed to the same cost-bounded rule-driven solver that closes ``Then \dots'' gaps; a statement whose obligations cannot be closed is rejected with a localized diagnostic. The rule format, the runtime, the well-definedness mechanism, and the asymmetric-context details all live in Appendix~\ref{sec:solver}.

\paragraph{Lean emission (optional).} For users who want external verification or to feed downstream Lean tooling, an accepted proof can be re-expressed as a Lean file. The formatter itself does no proof search, but the emitted Lean code calls Lean-side tactics --- \texttt{obvious}, \texttt{simp}, \texttt{assumption}, custom rewrite tactics --- on small per-step obligations, which \texttt{lake build} then closes. This stage is not required for Visored to accept or reject a proof.

\paragraph{What an LLM sees.} From the LLM's perspective, Visored is a function \emph{LaTeX proof $\to$ \{accepted, or a diagnostic pointing at a specific source location\}}, with Lean output available as a side product when wanted. Intermediate states are exposed as human-readable trace output rather than as a generic serialized format, so an LLM can use Visored both as an end-to-end checker and as a generator of training data; an obvious next step is to wire the diagnostic channel directly as an RL reward signal.

%% file: sections/arch_pipeline_fig.tex
\begin{figure}[t]
\centering
\begin{tikzpicture}[
  font=\small,
  >=Stealth,
  stage/.style={draw, rounded corners, align=center, minimum width=4.2cm, minimum height=0.78cm, fill=blue!4},
  io/.style={draw, align=center, minimum width=4.2cm, minimum height=0.78cm},
  fail/.style={draw, dashed, rounded corners, align=center, fill=red!4, inner sep=3pt},
  opt/.style={draw, dotted, rounded corners, align=center, fill=black!3, inner sep=3pt},
  lbl/.style={font=\scriptsize, midway},
  node distance=7.5mm and 12mm,
]
\node[io] (latex) {LaTeX source \;{\scriptsize(CNL inside \texttt{example})}};
\node[stage, below=of latex] (syn) {Syntax tree};
\node[stage, below=of syn] (sem) {Typed AST (sem)};
\node[stage, below=of sem] (mir) {Visored MIR};
\node[stage, below=of mir] (elab) {Elaboration {\scriptsize+ well-definedness}};
\node[stage, below=of elab] (solver) {Cost-bounded solver};
\node[draw, rounded corners, fill=green!8, align=center, minimum width=4.2cm, minimum height=0.78cm, below=of solver] (acc) {\textbf{accept} {\scriptsize(Visored verdict)}};

\node[fail, left=of solver] (diag) {localized\\diagnostic};
\node[opt, right=of acc] (uvl) {UVL MIR};
\node[opt, below=of uvl] (lean) {Lean file\\{\scriptsize\texttt{lake build}}};

\draw[->] (latex) -- node[lbl, right] {parse} (syn);
\draw[->] (syn) -- node[lbl, right] {types, kinds} (sem);
\draw[->] (sem) -- node[lbl, right] {lower} (mir);
\draw[->] (mir) -- (elab);
\draw[->] (elab) -- node[lbl, right] {discharge ``Then\ldots''} (solver);
\draw[->] (solver) -- (acc);

\draw[->, dashed] (solver) -- (diag);
\draw[->, dashed] (elab.west) to[out=180, in=90] (diag.north);

\draw[->, dotted] (acc) -- node[lbl, above] {transcribe} (uvl);
\draw[->, dotted] (uvl) -- (lean);
\end{tikzpicture}
\caption{The Visored pipeline. Each stage lowers the previous stage's output into a more constrained form or refuses with a localized diagnostic that points at the offending source span (dashed; any stage can refuse, not only the two drawn). Visored's verdict, accept or diagnostic, is produced by the solver and does not pass through Lean. Once a proof is accepted, the dotted branch optionally transcribes it, together with the recorded derivation steps, to UVL MIR and then to a Lean file for external kernel re-checking.}
\label{fig:pipeline}
\end{figure}
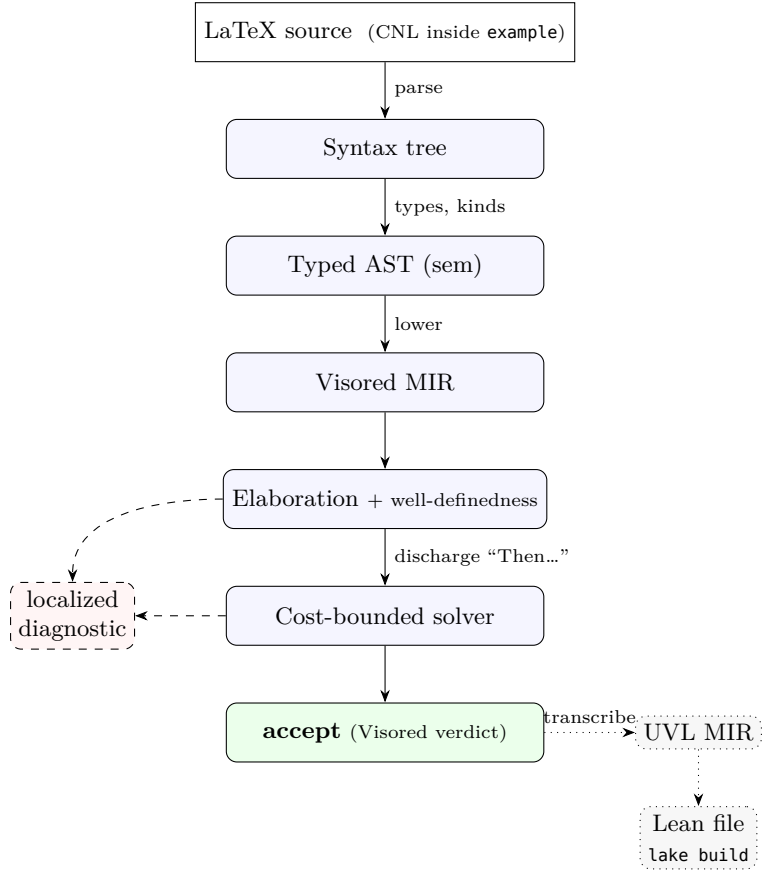

%% file: sections/worked-example.tex
\section{Worked Example}
\label{sec:example}

To make the value proposition concrete, the three examples below show the same kind of object written two ways: a Visored CNL input (left column) and the kind of Lean source an existing LLM autoformalization pipeline would have to produce directly (right column). The CNL stays close to what an LLM is already fluent at producing; the Lean additionally requires the LLM to pick the right Mathlib lemmas and tactics, which is the part of the pipeline where current direct-autoformalization systems report the largest pass-rate gap. All three pairs are checked automatically by the artifact \texttt{Makefile}: the CNL passes Visored and the Lean compiles against Mathlib.

%
  \par\addvspace{\medskipamount}%
  \noindent%
  \begin{minipage}[t]{0.47\textwidth}%
  \textbf{Visored CNL input (typeset)}\\[2pt]%
  \rule{\linewidth}{0.4pt}%
  {\setlength{\parskip}{0pt}\setlength{\parindent}{0pt}\linespread{0.9}\fontsize{7pt}{8pt}\selectfont\input{examples/_generated/ex01_cnl.tex}}%
  \end{minipage}\hfill%
  \begin{minipage}[t]{0.5\textwidth}%
  \textbf{Hand-written Lean 4 proof}\\[2pt]%
  \rule{\linewidth}{0.4pt}%
  \lstinputlisting[language=Lean4]{examples/_generated/ex01_proof.lean}%
  \end{minipage}%
  \par\smallskip%
  \noindent{\small miniF2F \texttt{mathd-algebra-116}. A two-line CNL substitution. The Lean has to handle \texttt{Real.sqrt 131}, the squaring lemma \texttt{Real.sq\_sqrt}, and a non-linear arithmetic discharge that needs $\sqrt{131}^2 = 131$ as an explicit hint.}\par\medskip%

\input{sections/worked-example-extra}

The full pipeline on the first input would proceed through the stages of Section~\ref{sec:architecture}: parse, type-check, elaborate the assumption and the goal, dispatch the goal to the solver, accept. An accepted proof can optionally be emitted as Lean source; the emitted Lean is more verbose than the hand-written version (Section~\ref{sec:limitations}) but is structurally a sequence of named lemmas, each provable by a small targeted tactic.

%% file: examples/_generated/ex01_cnl.tex
\begin{example}
Let $k\in\mathbb{R}$.

Let $x\in\mathbb{R}$.

Assume $x=\frac{13-\sqrt{131}}{4}$.

Assume $2x^2-13x+k=0$.

The goal is to prove $k=\frac{19}{4}$.

Then $k = 13x - 2x^2$.

Then $k = \frac{19}{4}$.
\end{example}

%% file: sections/worked-example-extra.tex

%
  \par\addvspace{\medskipamount}%
  \noindent%
  \begin{minipage}[t]{0.47\textwidth}%
  \textbf{Visored CNL input (typeset)}\\[2pt]%
  \rule{\linewidth}{0.4pt}%
  {\setlength{\parskip}{0pt}\setlength{\parindent}{0pt}\linespread{0.9}\fontsize{7pt}{8pt}\selectfont\input{examples/_generated/ex04_cnl.tex}}%
  \end{minipage}\hfill%
  \begin{minipage}[t]{0.5\textwidth}%
  \textbf{Hand-written Lean 4 proof}\\[2pt]%
  \rule{\linewidth}{0.4pt}%
  \lstinputlisting[language=Lean4]{examples/_generated/ex04_proof.lean}%
  \end{minipage}%
  \par\smallskip%
  \noindent{\small miniF2F \texttt{induction\_sum2kp1npqsqm1}. $\sum_{k=0}^{n-1} (2k+3) = (n+1)^2 - 1$. The CNL is three lines; the Lean side does explicit induction, \texttt{Finset.sum\_range\_succ}, and \texttt{omega} for truncated $\mathbb{N}$ subtraction.}\par\medskip%

  \par\addvspace{\medskipamount}%
  \noindent%
  \begin{minipage}[t]{0.47\textwidth}%
  \textbf{Visored CNL input (typeset)}\\[2pt]%
  \rule{\linewidth}{0.4pt}%
  {\setlength{\parskip}{0pt}\setlength{\parindent}{0pt}\linespread{0.9}\fontsize{7pt}{8pt}\selectfont\input{examples/_generated/ex09_cnl.tex}}%
  \end{minipage}\hfill%
  \begin{minipage}[t]{0.5\textwidth}%
  \textbf{Hand-written Lean 4 proof}\\[2pt]%
  \rule{\linewidth}{0.4pt}%
  \lstinputlisting[language=Lean4]{examples/_generated/ex09_proof.lean}%
  \end{minipage}%
  \par\smallskip%
  \noindent{\small miniF2F \texttt{mathd-algebra-462}. $(1/2+1/3)(1/2-1/3)=5/36$. Closed by \texttt{norm\_num}; an LLM that tries \texttt{ring} will fail because the constants are rational.}\par\medskip%

%% file: examples/_generated/ex04_cnl.tex
\begin{example}
Let $n\in\mathbb{N}$.

The goal is to prove $\sum_{k=0}^{n-1} (2k+3)={(n+1)}^2-1$.

We have $\sum_{k=0}^{n-1} (2k+3) = n^2 + 2n$.

We have ${(n+1)}^2 - 1 = n^2 + 2n$.

Then $\sum_{k=0}^{n-1} (2k+3)={(n+1)}^2-1$.
\end{example}

%% file: examples/_generated/ex09_cnl.tex
\begin{example}
The goal is to prove $\Bigl(\frac{1}{2}+\frac{1}{3}\Bigr)\Bigl(\frac{1}{2}-\frac{1}{3}\Bigr)=\frac{5}{36}$.

We have $\Bigl(\frac{1}{2}+\frac{1}{3}\Bigr)\Bigl(\frac{1}{2}-\frac{1}{3}\Bigr) = {\Bigl(\frac{1}{2}\Bigr)}^2 - {\Bigl(\frac{1}{3}\Bigr)}^2$.

We have ${\Bigl(\frac{1}{2}\Bigr)}^2 - {\Bigl(\frac{1}{3}\Bigr)}^2 = \frac{1}{4} - \frac{1}{9}$.

We have $\frac{1}{4} - \frac{1}{9} = \frac{5}{36}$.

Then $\Bigl(\frac{1}{2}+\frac{1}{3}\Bigr)\Bigl(\frac{1}{2}-\frac{1}{3}\Bigr)=\frac{5}{36}$.
\end{example}

%% file: sections/intended-usage.tex
\section{Intended Usage}
\label{sec:intended-usage}

Visored is designed to be useful in four modes inside an LLM workflow. Each addresses a different point at which an LLM workflow currently lacks reliable mathematical feedback. The four are design intent rather than measured outcomes in this first paper: Section~\ref{sec:experiments} reports a coverage study closest to the autoformalizer mode, and Section~\ref{sec:limitations} lists exercising the modes as future work. Throughout this section we use \emph{low-level prover} to mean a kernel-checked theorem prover with a code-like surface, such as Lean, Rocq, or Isabelle.

\paragraph{Inference-time verifier.} At deployment, Visored sits alongside the LLM as a real-time mathematical checker: the LLM emits a CNL draft, Visored either accepts it or returns a localized diagnostic, and the user sees only formally-checked output while the interaction surface stays in natural language. This is the deployment shape an end product would want --- a math chatbot, tutoring system, or research assistant that does not silently emit a hallucinated proof, without forcing the user to read formal-prover syntax to know whether the answer is correct.

\paragraph{Autoformalizer loop.} The LLM is prompted to write a LaTeX proof. Visored returns either accepted or a localized diagnostic. The LLM rewrites the offending region and retries. Visored's CNL surface stays close to natural mathematical writing, so anyone who knows math can inspect both the proof and the diagnostic. Compared to direct autoformalization targeting a low-level prover, the human-in-the-loop barrier could be lower: a research mathematician who reads CNL fluently but not the target prover can still sit in the loop and contribute to a formalization effort.

\paragraph{Dense reward for RL.} Each elaboration step in Visored produces a check that can succeed or fail independently. The per-step verdict can be exposed as a vector reward (one entry per step) rather than a single scalar at the end of an attempt. Because Visored strips away low-level prover details and stays hierarchical at the CNL level, the resulting signal is closer in spirit to a natural-language self-verifier than to a low-level syntax checker, and the CNL hierarchy matches the mathematical decomposition of the proof rather than the prover's tactic substrate.

\paragraph{CNL data generator.} A pipeline that prompts an LLM to write LaTeX proofs at scale and filters with Visored is in effect an infinite generator of formally correct mathematical proofs in a natural-language surface. The CNL vocabulary itself is configurable through spec files (Appendix~\ref{sec:syntax}) and can be tuned arbitrarily close to the conventions of any particular mathematical literature, so the same pipeline can produce competition-math style, textbook style, or research-paper style proofs as needed. The result is a free, unbounded corpus of formally checked yet naturally-reading mathematical proofs --- a renewable supply of clean math data that does not have to be curated by hand or scraped from existing texts. Compared to scraping or auto-generating text from a low-level prover, CNL data could give better transfer to natural-mathematical-writing tasks, because the surface stays close to the distribution an LLM is already fluent on.

%% file: sections/experiments.tex
\section{Experiments}
\label{sec:experiments}

We report a single, deliberately narrow result and do not oversell it. This first version of Visored is meant as a prototype: its purpose is to work out the fundamental design issues across the four layers it is built from --- the CNL \emph{syntax}, its \emph{semantics}, the \emph{solver}, and \emph{Lean transpilation} --- not to be a finished, competitive prover. It is consequently an older system that has accumulated substantial technical debt, and what follows is not an independent benchmark but a coverage study on one split of miniF2F~\cite{minif2f}, conducted while the system was still being built. The number should be read as a lower bound on what the design can already express, not as a tuned, competitive pass rate.

\paragraph{Setup.} We work on miniF2F-valid (244 problems). A single LLM coding agent (Claude, driving Visored through a documented skill that describes the CNL surface language and how to write and verify proofs in it) is handed each problem's formal statement as a CNL prefix and asked to finish the proof. The loop is the ordinary agentic one: the agent drafts proof steps, runs the Visored CLI, reads the diagnostic on failure, and revises. When a step exposed a gap --- a deduction a mathematician would consider routine but that Visored could not close --- we recorded it and, where feasible, added the corresponding rule to the library before continuing. The valid split therefore served as \emph{both} our development set and our evaluation set, and no bounded retry or sampling budget was imposed. This is exactly the kind of co-development that makes the result a statement about expressivity, not about held-out generalisation.

\paragraph{Coverage on miniF2F-valid.} Under this protocol the agent produced Visored-checked proofs for \textbf{222 of 244 problems (91\%)}. Table~\ref{tab:minif2f-valid} breaks this down by problem source. Coverage is near-saturated everywhere except the olympiad (IMO) problems, which are both the hardest mathematically and the least supported by the current rule database.

\begin{table}[h]
\centering
\begin{tabular}{lrr}
\toprule
Category & Solved / Total & Rate \\
\midrule
Induction                       &  8 / 8   & 100\% \\
Algebra (mathd + custom)        & 86 / 88  &  98\% \\
Number theory (mathd + custom)  & 66 / 68  &  97\% \\
AMC                             & 42 / 45  &  93\% \\
AIME                            & 13 / 15  &  87\% \\
IMO                             &  7 / 20  &  35\% \\
\midrule
Total                          & 222 / 244 & 91\% \\
\bottomrule
\end{tabular}
\caption{miniF2F-valid coverage by problem source. ``Solved'' means the agent produced a proof that Visored accepts; rules were added to the library as gaps were discovered, so these are coverage figures on the development split rather than held-out pass rates.}
\label{tab:minif2f-valid}
\end{table}

\paragraph{Where it fails.} Of the 22 unsolved problems, 13 are IMO problems. We want to be careful about attribution: these are almost entirely \emph{Visored} gaps, not the agent failing at the mathematics. In every case the obstacle is a feature, rule, type, or piece of machinery that Visored does not yet have --- a missing AM--GM inequality, a sum of $\lfloor \log \rfloor$ terms, finite-field types $\mathbb{Z}/n\mathbb{Z}$ that are unimplemented (two problems), limit reasoning, trigonometric identities and sums, and, for the olympiad problems, more substantial proof machinery (such as the inequality and descent techniques behind IMO 1988 P6 or IMO 1978 P5). We have no evidence that the agent could not drive these proofs once the machinery is in place --- we simply have not added it yet. One further failure (IMO 1987 P4) is a plain regression --- an existential assertion that used to check and no longer does --- which is the characteristic signature of accumulated technical debt. Table~\ref{tab:minif2f-unsolved} lists all 22 with the specific obstacle in each case; the broader scope limitations are discussed in Section~\ref{sec:limitations}.

\begin{table}[h]
\centering
\scriptsize
\begin{tabular}{l l p{7.3cm}}
\toprule
Problem & Source & Why it remains unsolved \\
\midrule
\multicolumn{3}{@{}l}{\textit{(a) Needs substantial mathematical machinery Visored does not yet have}}\\
\texttt{imo-2006-p6}  & IMO & Competition-level inequality. \\
\texttt{imo-1962-p4}  & IMO & Nonlinear trigonometric equation. \\
\texttt{imo-1978-p5}  & IMO & Rearrangement inequality. \\
\texttt{imo-1988-p6}  & IMO & Vieta jumping; a famously hard olympiad problem. \\
\texttt{imo\_1990\_p3} & IMO & Long divisibility-chain argument. \\
\addlinespace
\multicolumn{3}{@{}l}{\textit{(b) Genuine prover gaps --- a missing rule, capability, or type}}\\
\texttt{algebra-amgm-prod1toneq1-sum1tongeqn} & Algebra & General AM--GM inequality is not in the rule library. \\
\texttt{mathd\_algebra\_31}       & Algebra & Requires limit / fixed-point reasoning. \\
\texttt{aime-1994-p4}            & AIME & Evaluating a sum of $\lfloor \log \rfloor$ terms. \\
\texttt{aime\_1997\_p12}          & AIME & Closed form for a sum of sines and cosines. \\
\texttt{amc12b-2021-p21}         & AMC  & Transcendental equation (the agent also failed to find a proof). \\
\texttt{amc12a\_2003\_p25}        & AMC  & Reasoning about a set fixed point $S=f(S)$. \\
\texttt{amc12a\_2003\_p24}        & AMC  & Set maximisation / optimisation. \\
\texttt{mathd-numbertheory-232}  & NT   & Finite-field type $\mathbb{Z}/31\mathbb{Z}$ not supported. \\
\texttt{mathd\_numbertheory\_668} & NT   & $\mathbb{Z}/n\mathbb{Z}$ constructor unimplemented. \\
\texttt{imo\_1966\_p4}            & IMO  & Needs trigonometric identities ($\tan = \sin/\cos$ not known). \\
\texttt{imo\_1966\_p5}            & IMO  & System of absolute-value equations. \\
\texttt{imo\_1967\_p3}            & IMO  & Product divisibility. \\
\texttt{imo\_1977\_p5}            & IMO  & Disjunction in the conclusion (division with remainder). \\
\texttt{imo\_1979\_p1}            & IMO  & Alternating harmonic sum / number theory. \\
\texttt{imo\_1987\_p6}            & IMO  & Prime-generating polynomial argument. \\
\texttt{imo\_1993\_p5}            & IMO  & Existence of a function (construction). \\
\addlinespace
\multicolumn{3}{@{}l}{\textit{(c) Regression}}\\
\texttt{imo-1987-p4}  & IMO & A bare existential assertion that used to check no longer does. \\
\bottomrule
\end{tabular}
\caption{The 22 miniF2F-valid problems Visored does not solve, grouped by what they need: (a) substantial mathematical machinery not yet built, (b) a specific missing rule, capability, or type, (c) a regression. All three are Visored-side gaps; none is attributed to the agent being unable to do the mathematics.}
\label{tab:minif2f-unsolved}
\end{table}

\paragraph{The cost of Lean transpilation, and why we stopped here.} The gaps above persist not because they are conceptually hard to close but because of where this prototype spent its effort. The dominant sink was Lean transpilation: we spent several months on it, largely because we had not yet found the right way to organize the emitted proofs. The lasting consequences are a very large Lean output\footnote{Representative excerpts of this emitted Lean---the first chunks of one accepted proof per miniF2F subject, each annotated with its full size---are
  public at \url{https://github.com/xiyuzhai-husky-lang/visored}.}--- Section~\ref{sec:limitations} reports a median of $\approx 250\times$ the hand-written proof, with a tail to $\sim\!25{,}000\times$ (see Appendix~\ref{sec:transpilation}) --- and an incremental Lean recompilation far heavier than the design requires. The effect is best stated comparatively: iterating on this codebase --- add a rule, regenerate, recompile --- is much slower than the same loop in the successor prover we are now building with the design lessons this prototype taught us. So rather than keep grinding features into the old system, we stopped its development once it had served its purpose: showing us how the four layers --- syntax, semantics, the solver, and Lean transpilation --- should be organized. The unsolved problems are therefore a backlog we chose not to clear on this codebase, not a frontier of what the design can express.

\paragraph{How to read these numbers.} The honest claim is modest: Visored's CNL surface and its current rule library are already expressive enough that an LLM agent, in an ordinary verify-and-revise loop, can drive 91\% of miniF2F-valid to a checked proof. We do not claim competitive automated theorem proving, we do not extrapolate to the test split, and we do not compare against systems run under different protocols. We stop here deliberately.



%% file: sections/limitations.tex
\section{Limitations and Future Work}
\label{sec:limitations}

We state current limitations honestly because they are concrete and addressable, not foundational.

\paragraph{Math scope.} The current rule databases and prelude cover arithmetic, basic algebra, set theory, elementary number theory, and parts of real analysis. Categories of miniF2F that require, e.g., heavy combinatorial enumeration or sophisticated inequality manipulation are currently out of scope; supporting them is a matter of adding rules, not changing the foundation.

\paragraph{Emitted-Lean verbosity.} The current sparse emission scheme is extremely verbose --- per-problem line-count ratios on miniF2F-valid are heavy-tailed, with median $\approx 250\times$ and a tail reaching $\sim 25{,}000\times$. The full statistics and the design choices that drive them are in Appendix~\ref{sec:transpilation}.

\paragraph{Solver diagnostics.} When the Miracle budget is exhausted without closing a goal, the diagnostic reports the unclosed goal and source location but does not yet report which rule almost fired or how close the best candidate was. Surfacing this would tighten the LLM-in-the-loop signal.

\paragraph{Type-system ceiling.} The dependent-type system is intentionally lightweight. Mathematics whose statement requires non-trivial dependent indexing or higher inductives is outside the current scope. Visored's role here is complementary: produce a usable formal corpus for elementary mathematics, and hand off harder cases to the host prover.

\paragraph{Engineering maturity.} Visored is research code. Rule organization, spec file layout, and IR boundaries are still moving; the system is a checkpoint of an evolving design rather than a stable platform.

\paragraph{Exercising the intended-usage modes.} Each of the four modes in Section~\ref{sec:intended-usage} is currently a design intent rather than a measured outcome. A natural next step is to actually instantiate each: deploy Visored as an inference-time verifier behind an LLM endpoint, drive an autoformalizer loop on miniF2F with a frozen LLM, train an RL policy using Visored's per-step verdict as a dense reward, and stand up the CNL-data-generation pipeline at scale.

\paragraph{Richer CNL.} The current CNL vocabulary covers the constructions needed for miniF2F-style proofs but is still terse compared to textbook prose. Adding more idiomatic patterns --- longer sentence templates and standard shorthands like ``for sufficiently large $n$'', ``WLOG'', ``passing to a subsequence'' --- would make the CNL strictly more readable and closer to research-paper conventions.

\paragraph{Tactic mode through hints.} The current solver runs autonomously: the user writes ``Then \dots'' and Visored either closes the gap or fails. A tactic mode in which the user supplies optional hints (``by induction on $n$'', ``apply Cauchy-Schwarz'', ``rewrite using lemma $X$'') would let an LLM (or a human) steer the solver through hard steps without leaving the CNL surface.

%% file: appendix/related-work.tex
\section{Extended Related Work}
\label{app:related-work}

\subsection{Progress of AI for Mathematics}
\label{sec:rw-ai-math}
\input{sections/related-work/ai-math}

\subsection{Prover Designs}
\label{sec:rw-provers}

\input{sections/related-work/intro}

\paragraph{Direct LLM autoformalization and whole-proof Lean models.} The natural baseline is to fine-tune or prompt an LLM to map an informal statement and/or proof directly to Lean or Isabelle source. Wu et al.~\cite{wu2022autoformalization} establish the basic setup for Isabelle/HOL and report state-of-the-art on miniF2F at the time. Draft, Sketch, and Prove~\cite{jiang2023dsp} extends this by first drafting an informal proof, sketching a formal skeleton, and letting an automated theorem prover fill the gaps. A fast-moving line of whole-proof Lean models has since pushed pass rates much higher: DeepSeek-Prover and its V2~\cite{xin2024deepseekprover,ren2025deepseekproverv2} scale synthetic data and reinforcement learning targeting Lean 4; STP~\cite{dong2025stp} trains a conjecturer and a prover against each other to escape the limited supply of formal data; Goedel-Prover and its V2~\cite{lin2025goedelprover,lin2025goedelproverv2} add scaffolded data synthesis and verifier-guided self-correction; Kimina-Prover~\cite{wang2025kimina} trains a large formal reasoning model with long-form reasoning; and Seed-Prover~\cite{chen2025seedprover} adopts lemma-style proving, reporting that it saturates miniF2F while solving most recent IMO problems. Two things separate this line from Visored. First, these systems take an \emph{already formalized} statement as input and produce a proof, whereas Visored's input is the informal LaTeX surface; the hard surface-to-semantics decisions sit upstream of where these provers begin. Second, the natural checkpoint of an end-to-end pipeline is the final Lean file as a whole, so per-attempt feedback to the LLM is typically a single accept/reject after compilation. Visored instead keeps a checkable IR at every stage, so failures are localized to a specific sub-expression and the per-stage diagnostic is itself usable as a denser reward signal. This is a design difference, not a critique: end-to-end translation and stage-by-stage elaboration are complementary points in the design space.

\input{sections/related-work/cnl}

\input{sections/related-work/tactic-layers}

\input{sections/related-work/proof-checkers}

\input{sections/related-work/well-definedness}

\input{sections/related-work/informal-aligned}

\input{sections/related-work/synthesis}

%% file: sections/related-work/ai-math.tex

LLMs built on the Transformer~\cite{vaswani2017attention} have become capable mathematical reasoners, with steady progress on benchmarks such as GSM8K~\cite{cobbe2021gsm8k}, MATH~\cite{hendrycks2021math}, and miniF2F~\cite{minif2f}, and further gains from math-specialized pretraining as in Minerva~\cite{minerva} and Llemma~\cite{llemma}. At the top of competition mathematics, two results from the same lab one year apart are telling: AlphaProof reached IMO 2024 silver by writing \emph{formal} (Lean) proofs via reinforcement learning (RL)~\cite{alphaproof2024}, while a Gemini Deep Think configuration reached IMO 2025 gold writing proofs in \emph{natural language}, graded like human contestants~\cite{gemini-imo-gold}. That the higher medal went to the informal-reasoning system is part of why we keep a prover's surface close to natural language.

Beyond competition mathematics, AI has begun to contribute to research-level open problems. AlphaEvolve improved or matched the state of the art on dozens of them, including Erd\H{o}s's minimum-overlap conjecture and kissing-number bounds in dimension 11~\cite{alphaevolve2025}; three Erd\H{o}s problems fell to AI-assisted constructions in one week of late 2025, each verified by Terence Tao~\cite{tao2025erdosstreak}; and an OpenAI reasoning model beat the long-conjectured grid optimum for the 1946 Erd\H{o}s planar unit-distance problem~\cite{openai2026unitdistance}. Smaller academic efforts follow the same pattern: ThetaEvolve improves best-known bounds on circle packing and an auto-correlation inequality with an 8B open model~\cite{wang2025thetaevolve}, a follow-on autonomous-agent scaffold tightens the Ramsey number $R(3,17)$ for the first time in decades and lifts $R(4,15)$ past the AlphaEvolve record~\cite{wang2026ramsey}, and LLM-generated geometric lemmas raise the certified lower bound for the Steiner ratio, the Gilbert-Pollak conjecture~\cite{ke2026gilbertpollak}. Tao~\cite{tao2025exploration,tao2025notices,tao2026openaiacademy} reads this shift as moving the binding constraint from generating mathematical ideas to verifying and organizing them at scale. That is exactly what Visored targets: not another proof-search engine, but infrastructure for checking and organizing AI-produced informal mathematics.

%% file: sections/related-work/intro.tex
We position Visored against four families of work an AI audience is likely to know: (i) direct LLM-based autoformalization, (ii) controlled-natural-language (CNL) provers, (iii) human-friendly natural-language tactic layers on top of existing provers, and (iv) rule-based proof checkers for structured natural-language proofs.

%% file: sections/related-work/cnl.tex
\paragraph{CNL provers.} Mizar~\cite{mizar} has, since the 1970s, accepted proofs written in a mathematical vernacular --- a controlled English-like syntax compiled against the Mizar Mathematical Library; the longevity of the project is evidence that mathematically structured natural language can carry a precise formal semantics. Isabelle/Naproche~\cite{naproche2021} is the modern reference point: a ForTheL CNL frontend whose obligations are discharged by automated theorem provers via TPTP. Differences with Visored:
\begin{itemize}
\item \textbf{Surface coverage.} Both Naproche and Visored accept a controlled subset of mathematical English, not unrestricted prose. The difference is in how the subset is defined and extended. ForTheL is a fixed, explicitly designed grammar that the writer learns; widening it means a Naproche release. Visored's accepted subset is defined externally by \texttt{.lpcsv} spec files (Appendix~\ref{sec:syntax}) and extended by adding entries rather than by modifying the parser. The default vocabulary aims to cover what a typical LLM emits when asked to write a proof in LaTeX, but inputs outside the configured subset are rejected, with a localized diagnostic, just as in Naproche.
\item \textbf{Foundation.} Naproche translates ForTheL into first-order logic. Visored is dependently typed, which is what lets the same proof be re-expressed in a dependently typed target like Lean 4.
\item \textbf{Solver philosophy.} Naproche calls out to general-purpose ATPs (E, Vampire) for each obligation. Visored's solver is custom and rule-database-driven, with a non-deterministic cost-bounded runtime (\emph{Miracle}); this trades general-purpose power for predictable per-step cost and per-step diagnostics, both of which matter for LLM-in-the-loop use.
\end{itemize}
GFLean~\cite{gflean2024} uses the Grammatical Framework as the parsing layer and emits Lean from a subset of ForTheL (``Simplified ForTheL''). It shares the CNL flavor but, by design, accepts a smaller surface language than Naproche or Visored, and is not aimed at an LLM-driven workflow.

Recent work has begun to use these CNL systems as the translation interface for autoformalization rather than the end target. De Lon et al.~\cite{lemiz2025} present a bidirectional verifiability-preserving syntactic translation between Mizar and a Naproche-ZF--style CNL, with the explicit motivation that LLMs have seen far more natural-language mathematics in training than formal source. The EuroProofNet WG5 white paper~\cite{europroofnetcnl2025} surveys the integration of CNL in formal mathematics systems and articulates the shared design space (informalization, synthetic-data pipelines, ATP back-ends). Both treat CNL as a layer wrapped around an existing formal system. Visored shares the CNL premise but inverts the role: CNL is the primary medium the prover operates on, and existing formal systems are reached --- if at all --- as optional emission targets.

%% file: sections/related-work/tactic-layers.tex
\paragraph{Natural-language tactic layers on top of existing provers.} Verbose Lean~\cite{verboselean} provides Lean 4 tactics whose surface syntax reads as controlled English (or French), targeted at teaching pen-and-paper proof writing. Waterproof~\cite{waterproof} plays the analogous role for Rocq/Coq, with English-language tactic wrappers and an automation tactic (\texttt{waterprove}) over customizable hint databases, again aimed at undergraduate teaching. Both are UX layers rather than independent provers: the user is still writing in (a more readable form of) the host prover, and the type system, library, and verifier are all the host's. Visored sits earlier in the stack --- the user writes mathematical English / LaTeX, and the host prover is reached only as an emission step, if at all.

%% file: sections/related-work/proof-checkers.tex
\paragraph{Proof checkers.} ProofGrader~\cite{proofgrader} is a rule-based checker for natural-language-like proofs, with its own formal semantics for each proof construct. The ProofGrader project's focus, as we read it, was not on building or targeting a minimal foundational kernel but on the surface syntax and solver design that together give a natural API for textbook-style mathematical proofs --- a design goal closely aligned with Visored's. The ProofGrader solver --- a configurable solver manager with per-solver cost budgets, dynamic priority scheduling, and hierarchical sub-proposition decomposition --- fits the structure of textbook mathematical proofs better than SAT/SMT, and is the closest design match to Visored's Miracle runtime. Visored complements this line of work with a dependently-typed IR, and accepted proofs can be optionally re-emitted as Lean 4 source for users who want external kernel-level re-verification.

%% file: sections/related-work/well-definedness.tex
\paragraph{Well-definedness at elaboration time.} Whether $1/x$ should be accepted before $x \neq 0$ has been established splits the landscape. PVS~\cite{pvs1992,pvs-subtypes} attaches the side conditions to the \emph{types} of partial functions via predicate subtyping --- division has signature $[\text{real},\text{nonzero\_real}\to\text{real}]$, square root has $[\text{nnreal}\to\text{nnreal}]$, log has $[\text{posreal}\to\text{real}]$ --- so applying any of them to a wider-typed argument generates a \emph{type-correctness condition} (TCC) at typecheck time. The dominant alternative --- Lean, Rocq, Mathlib, Mizar --- makes the partial functions \emph{total} by convention ($1/0 \equiv 0$ in Lean and Rocq~\cite{xena-divbyzero}; Mizar follows the same convention) and surfaces the side condition only when a downstream lemma needs it. The landscape factors along two axes:

\begin{center}
\small
\begin{tabular}{lll}
\toprule
 & \textbf{Surface} & \textbf{Well-definedness mechanism} \\
\midrule
PVS                                       & code-like (spec language) & predicate subtyping $\to$ TCCs at typecheck \\
Lean / Rocq / Mizar / Mathlib             & code-like                 & totalization convention ($1/0 \equiv 0$) \\
Visored                                   & CNL (LaTeX)               & implicit hypothesis arguments per expression \\
\bottomrule
\end{tabular}
\end{center}

Visored takes the upfront-discipline path of PVS, but realizes it by treating every user-written expression as possibly carrying implicit hypothesis arguments that the elaborator must discharge as proof obligations, rather than by encoding them into a richer type system. The obligations are discharged on the spot by the same solver that closes ``Then \dots'' gaps; see Appendix~\ref{sec:solver} for the mechanism. The pay-off for an LLM workflow is a single uniform diagnostic stream covering both missing lemmas and missing positivity / non-zero conditions, with no transition to a separate specification language.

%% file: sections/related-work/informal-aligned.tex
\paragraph{Natural formalized languages.} A recent position paper by Cao, Xie, and Yan~\cite{cao2025nlaligned} argues that the LLM era calls for a new class of theorem provers whose surface is a \emph{natural formalized language} --- a proof language with three required properties: (i) rigorous formal semantics, (ii) maximal approximation to how mathematicians actually write, and (iii) accurate representation of the proof structures that conventional provers were not designed to express. The structures the paper highlights as out of reach for traditional provers include instantiation indicated by identical variable names, the fluid free-or-bound status of a variable across a proof, the gap between definite and indefinite binders in integral notation, and the postulate--derive--verify pattern of problem-solving proofs. The paper does not propose a specific implementation; it identifies these informal-proof features and calls for the design space to be explored. Visored is one concrete attempt at a prover in this class: the CNL surface (Appendix~\ref{sec:syntax}), the spec-driven extensible vocabulary, the implicit-hypothesis-argument treatment of partial functions (Appendix~\ref{sec:semantics}), the per-step diagnostic channel, and the rule-driven solver are all attempts at the design moves the position paper sketches.

%% file: sections/related-work/synthesis.tex
\paragraph{Visored as an attempted synthesis.} Read together, the four families above describe a design space that has been partially explored from four different directions. Each direction contributed something distinct: Mizar and Naproche showed that mathematically structured natural language can carry a precise formal semantics; PVS showed that partial-function side conditions belong at typecheck time, with context flowing asymmetrically through boolean connectives; Lean / Rocq and their Mathlib-style libraries showed that a dependently-typed kernel plus a large rule library is a workable foundation for modern formal mathematics; ProofGrader and related rule-based checkers showed that cost-budgeted, priority-scheduled search over a domain-specific rule database fits the structure of textbook proofs better than SAT/SMT engines; and the recent LLM-autoformalization literature reframed the goal as producing usable training and reward signal for foundation models rather than just verified artifacts.

Visored is best understood as an early-prototype attempt to combine these into one stack: a LaTeX/CNL surface (Mizar / Naproche tradition), a dependently-typed IR with predicate-subtyping-style well-definedness (PVS), a rule-database-driven cost-bounded solver (the ProofGrader school), optional emission to a dependently-typed host prover for kernel-level re-verification (Lean), and a localized, structured diagnostic channel designed from the start to be consumed by an LLM in a retry / RL loop (the autoformalization literature). We do not claim any individual ingredient is novel; what we are testing is whether the combination is more than the sum of its parts for an AI-driven workflow. The experiments in Section~\ref{sec:experiments} and the limitations in Section~\ref{sec:limitations} report where the combination works today and where it does not.

%% file: appendix/syntax.tex
\section{Syntax}
\label{sec:syntax}

This appendix describes the surface language and the parser that turns it into a Visored syntax tree. The vocabulary --- both the sentence templates and the LaTeX commands they wrap --- lives in external \texttt{.lpcsv} (lisp-csv) spec files. Extending Visored to recognize a new sentence form or a new LaTeX command is a matter of adding entries to those files; the parser is not touched.

\subsection{Sentence templates}
\label{sec:syntax-sentence}

Proof content lives inside \texttt{example} environments and is structured into sentences. The sentence-template vocabulary maps natural-language patterns to abstract syntactic constructors. Representative entries:

\begin{quote}\small
\begin{verbatim}
`Let {formula}`                                       => let
`Let {x: formula} be a function defined by {def}`     => let_function_like_defined_by
`Let {formula} be such that {prop}`                   => let_such
`Assume {formula}`                                    => assume
`Then {prop}`                                         => have
`Thus, {prop}`                                        => have
`Hence, {prop}`                                       => have
`Therefore, {prop}`                                   => have
`We have {prop}`                                      => have
`Since {p}, {q}`                                      => have_hinted q (since p)
`Note that {prop}`                                    => have
`The goal is to prove {prop}`                         => goal
`It is enough to prove {prop}`                        => show
`We prove {prop} by induction on {var}`               => induction
`We prove {prop} by contradiction`                    => apagoge
`We prove {prop} by working through different cases`  => cases
`We prove {prop} by arguing from both sides`          => bilateral
`Case {prop}`                                         => case
`Contradiction`                                       => contradiction
`Claim: {prop}`                                       => claim
\end{verbatim}
\end{quote}

Each constructor admits several natural-language variants, so common rewordings (``Thus'', ``Hence'', ``Therefore'') reach the same parse without per-variant special cases. The pattern grammar supports optional tokens, optional letters inside words (\texttt{view[s]} matches both \texttt{view} and \texttt{views}), and choice (\texttt{a|b}).

\subsection{Notional propositions}
\label{sec:syntax-notion}

``$X$ is $P$'' constructions are recognized as predicate applications, with positive and negated natural-language forms declared together so that \emph{``$X$ is non-empty''}, \emph{``$X$ is not empty''}, and \emph{``$X$ is nonempty''} all parse to $\neg \mathrm{empty}(X)$. The current vocabulary covers
\begin{quote}
\texttt{prime}, \texttt{even}, \texttt{odd}, \texttt{empty}, \texttt{finite}, \texttt{infinite}, \texttt{injective}, \texttt{surjective}, \texttt{bijective},
\end{quote}
each in several surface variants (``is finite'', ``is a finite set'', ``is an injection'', ``is an injective function'', ``is not an injection'', \dots).

\subsection{Existential introductions}
\label{sec:syntax-exists}

Existential introduction is one of the constructs where natural mathematical prose is the most varied. The same constructor is reached by all of the following surface forms (and more):

\begin{quote}\small
\begin{verbatim}
`There exists {x} such that {p}`           => let_such
`There exist {x} such that {p}`            => let_such
`Take {x} such that {p}`                   => let_such
`Choose {x} such that {p}`                 => let_such
`Assume there exists {x} so that {p}`      => let_such
`There exists {ty} {x} such that {p}`      => let_indefinite_such
`Let {x} be {ty} such that {p}`            => let_definite_such
\end{verbatim}
\end{quote}

Both ``be so that'' and ``be such that'' phrasings are accepted; both singular and plural existence (``there exists / there exist'') reach the same constructor.

\subsection{LaTeX math vocabulary}
\label{sec:syntax-math}

Inside the formula slot of any template, the user writes ordinary LaTeX math. Each LaTeX command is matched against an entry in the math-vocabulary spec files, which pin down its arity, fixity, and meaning. The base vocabulary covers:

\begin{itemize}
\item \textbf{Number systems and propositional types:} $\mathbb{N}, \mathbb{Z}, \mathbb{Q}, \mathbb{R}, \mathbb{C}$, $\mathsf{Prop}$, $\mathsf{True}$, $\mathsf{False}$.
\item \textbf{Constants:} $e$, $\pi$, $i$.
\item \textbf{Comparison and equivalence:} $=, \ne, <, \le, >, \ge, \equiv, \approx$.
\item \textbf{Set relations and operations:} $\in, \notin, \subseteq, \supseteq, \subsetneq, \cup, \cap, \setminus$, $\bigcup, \bigcap$.
\item \textbf{Logical connectives and quantifiers:} $\land, \lor, \neg, \to, \leftrightarrow, \forall, \exists$.
\item \textbf{Arithmetic:} $+, -, \cdot, \times, /, \frac{}{}$, $\pm$, $\bmod$, $\pmod{}$.
\item \textbf{Number theory:} $\mid, \nmid$, $\gcd$, $\mathrm{lcm}$, $\binom{}{}$.
\item \textbf{Big operators:} $\sum, \prod, \int$, $\lim, \sup, \inf, \max, \min$.
\item \textbf{Common functions:} $\sin, \cos, \tan, \log, \ln, \exp, \sqrt{\,}$.
\item \textbf{Greek letters:} all lowercase (including \texttt{varepsilon}, \texttt{vartheta}, \texttt{varphi} etc.) and uppercase ($\Gamma, \Delta, \Theta, \Lambda, \Xi, \Pi, \Sigma, \Upsilon, \Phi, \Psi, \Omega$).
\end{itemize}

Subject-specific commands (calculus, complex analysis, elementary number theory) live in separate spec files and are loaded on demand. A LaTeX command the spec does not declare yields an \texttt{UnknownCommand} diagnostic pointing at the offending token.

\subsection{Environments}
\label{sec:syntax-env}

Document-structure environments are recognized but mostly transparent: \texttt{example}, \texttt{proof}, \texttt{theorem}, \texttt{lemma}, \texttt{corollary}, \texttt{proposition}, \texttt{equation}, \texttt{align}, \texttt{matrix}, \texttt{pmatrix}, \texttt{cases}, \texttt{itemize}, \texttt{enumerate}, \texttt{figure}, \texttt{table}. Proof content must live inside a proof-bearing environment such as \texttt{example}; structural environments outside that scope are ignored.

\subsection{From source to syntax tree}
\label{sec:syntax-tree}

Parsing produces a tree of \emph{Visored Syn} nodes. Each node carries the source span it came from, the resolved template or LaTeX command, and the (still untyped) sub-trees. This is the input to the Sem stage (Appendix~\ref{sec:semantics}). As described in Section~\ref{sec:architecture}, math mode and text mode are parsed with different strategies --- a precedence stack-based parser and a trie-like sentence-pattern matcher --- and both produce nodes in the same Syn tree.

\subsection{Scope and disambiguation}
\label{sec:syntax-scope}

Visored accepts a controlled subset of mathematical English embedded in LaTeX, not arbitrary prose. Within that subset the parser is intentionally tolerant of surface variation --- punctuation, whitespace, common synonyms for the same construct --- because these are cheap to enumerate in the spec. Inputs outside the configured subset are rejected with a localized diagnostic; we do not silently coerce or paper over them. The subset can be widened by adding spec entries; we have not had to modify the parser to support a new sentence form. Disambiguation that requires types --- e.g.\ deciding whether \verb|\,| in \texttt{f(x)\,y} is multiplication or function application --- is deferred to the Sem stage (Appendix~\ref{sec:semantics}) rather than overloading the parser.

%% file: appendix/semantics.tex
\section{Semantics}
\label{sec:semantics}

This appendix describes how a Visored syntax tree is given meaning --- the type and kind system, name resolution, and where well-definedness obligations enter the pipeline.

\paragraph{Kinds and types.} Every Visored expression carries a \emph{kind} (the broad ontological category --- \texttt{Num}, \texttt{Set}, \texttt{Prop}, \texttt{Function}, \dots) and a \emph{type} (the specific mathematical object --- $\mathbb{N}$, $\mathbb{R}$, $\mathbb{C}$, a function space, a finite set). Kinds are coarse and decidable; types are finer, dependent-flavored, and admit predicate subtyping (Appendix~\ref{app:related-work}). Atoms like $\sqrt{x}$ have a natural-domain type (\texttt{nnreal} for the principal real square root) and the elaborator emits a well-definedness obligation when the argument's type does not already match.

\paragraph{Bottom-up typing vs ML-style top-down elaboration.} A central design choice is that each subexpression in Visored is typed at the \emph{least type it naturally lives in}, rather than at whatever ambient type the surrounding context dictates. Contrast this with ML-style top-down elaboration (``ML'' here is the typed functional-language tradition, such as Standard ML and OCaml, not machine learning) as found in Lean / Rocq / Mathlib: when Lean elaborates a surface expression like $(3 : \mathbb{N}) \cdot x$ in an ambient type $\mathbb{C}$ with $x : \mathbb{R}$, it inserts coercions wherever they are locally needed --- on individual atoms or in unification-driven positions --- and the same mathematical expression can elaborate into different ASTs depending on where annotations and context sit. Visored builds expressions bottom-up from atoms: the multiplication $3 \cdot x$ first lives in $\mathbb{R}$ (the least common type of $\mathbb{N}$ and $\mathbb{R}$), and a coercion into $\mathbb{C}$ appears only at the outermost point where the type genuinely changes. The AST is canonical with respect to mathematical structure rather than annotation position, which is what makes the typed pattern matching used by the rule databases (Appendix~\ref{sec:solver}) tractable. When an accepted proof is re-expressed as Lean, a dedicated \texttt{pull\_cast\_rewrite} pass lifts the Lean-side ML-style coercion arrangement back into Visored's bottom-up canonical form before any Visored-derived lemma is applied.

\paragraph{Name resolution.} Each identifier is resolved against a stack of namespaces: the local proof context first (variables introduced by ``Let \dots'', ``Assume \dots''), then any \texttt{Set}-builder or quantifier binders currently in scope, then the global symbol table loaded from \texttt{.lpcsv} spec files. The resolver is conservative: an ambiguous identifier yields a diagnostic rather than a silent choice, so the LLM gets a localized signal it can act on.

\paragraph{Signature dispatch tables.} The semantic resolution of an operator application --- which signature to instantiate for $2 + x$ when $2 : \mathbb{N}$ and $x : \mathbb{R}$ --- is decided by a table lookup, not by a hand-written dispatcher. For each base operator class (binary operators, prefix operators, fractions, powers, square roots, etc.) a \texttt{.lpcsv} dispatch file declares rows of the form \texttt{(lhs-type, operator, rhs-type) $\to$ signature-ident}, and a separate \texttt{signature\_table.lpcsv} declares each signature's instantiation and the underlying signature-data variant. The \texttt{.lpcsv} format is a small Lisp dialect (lisp-csv) that supports parenthesized compound atoms like \texttt{(subset nat)}, which lets a single signature dispatch row cover an entire family of types uniformly. Adding a new operator overload or a new pair of numeric-type combinations is a matter of adding entries to these files; the elaborator source is not touched.

\paragraph{Sem AST.} The output of the Sem stage is a typed syntax tree (the \emph{sem AST}). Each node has a resolved type, a kind, the list of free variables it depends on, and the list of well-definedness obligations its sub-expressions generated --- these obligations are propagated upward with the context-flow rule described in Appendix~\ref{app:related-work} (under $a \land e$ the obligations of $e$ are guarded by $a$; under \emph{if-then-else} branches are guarded by the test and its negation).

\paragraph{Substrate for later stages.} The Sem AST is the substrate on which the MIR lowering, the solver, and the optional Lean emitter all operate. Pattern matching used by the rule databases (Appendix~\ref{sec:solver}) is typed pattern matching on this tree, not on raw text or the surface CNL.

%% file: appendix/kernel.tex
\section{Kernel: the Super Computation Graph}
\label{sec:kernel}

Visored's expressions, hypotheses, and derivations are all built on top of a single underlying graph structure --- the \emph{super computation graph} --- carried over from the first author's PhD thesis on deep learning theory~\cite{zhai2024thesis}, where the same structure was developed to reason about computational dependencies in deep networks. The proof-world analogue treats every object the elaborator manipulates as a node in this graph, and the graph's structure is what drives lemma parameter lists in Lean emission, hypothesis activation under quantifier and implication introduction, and the dispatch logic that survives changes in the active context.

\subsection{Reishi and reiatsu}
\label{sec:kernel-reishi}

A \emph{reishi} (霊子, ``spirit particle'') is a free node in the super computation graph. Each reishi carries a unique ID, a data variant (universe, variable, hypothesis, \dots), and recursively the reiatsu of its own upstream dependencies. Reishi are interned: the same logical dependency yields the same ID, which gives reishi-DAGs structural equality and lets the elaborator deduplicate dependencies across the proof.

A \emph{reiatsu} (霊圧, ``spiritual pressure'') is an ordered set of reishi: the transitive closure of a node's free-node dependencies. When a hypothesis or derivation is emitted as a Lean \texttt{theorem} or \texttt{lemma}, its reiatsu becomes that theorem's parameter list, and the reishi-DAG is walked in topological order so that every dependency is defined before use. This is the source of the parameter packets one sees on every emitted lemma (the ``coercion-witness arguments'' of Appendix~\ref{sec:transpilation} are concrete instances of reishi).

\subsection{Ascension}
\label{sec:kernel-ascension}

The super computation graph allows free nodes to be turned on and off: a downstream node is available only when its upstream free reishi are active. We call the operation of activating a previously-inactive free node and lifting a downstream proof through it \emph{ascension}.

Concretely, ascension lifts a node in the super computation graph --- any node, not only a proof; a term, a hypothesis, or a derivation step is equally a valid argument --- over an \emph{ascension step}: a $\textsf{Let}$ binder for a single variable, an $\textsf{Assume}$ for a hypothesis, or a forall / implication introduction. Several ascension steps can be chained.

The non-trivial part is that ascension must respect type dependency. Given $(a : A)$ and $(b : B\,a)$ with $b$'s type depending on $a$, ascending $a$ in isolation is ill-formed --- the downstream node $b$ refers to $a$, so $a$ has to be ascended together with $b$, and with anything else downstream of $a$ in the reishi-DAG. In general, an ascension step lifts a node together with the entire forward cone of its dependants. This is what lets the rest of the system manipulate proofs and terms uniformly at the level of bound variables and free assumptions without re-running the underlying reasoning, and it is the kernel-level operation behind the introduction of every universal quantifier and implication in an accepted Visored proof.

\subsection{Montage}
\label{sec:kernel-montage}

Because free nodes can be turned on and off, a downstream construction may become temporarily unavailable when an upstream free reishi is deactivated. A \emph{montage} is a small data structure that holds a list of objects each tagged with its reiatsu, and on query returns the first object whose reiatsu is currently alive. Montage is what lets the solver dispatch over multiple possible witnesses without crashing when some are temporarily out of scope --- a routine occurrence when ascension steps move proofs across activation boundaries.

\subsection{What this buys}
\label{sec:kernel-payoff}

The super computation graph as the underlying object has three concrete payoffs in Visored. First, hypothesis and derivation dependencies are tracked explicitly and globally, so a Lean lemma can be emitted with exactly the parameters it needs --- no more, no less --- without per-emission analysis. Second, ascension gives a uniform kernel-level account of quantifier and implication introduction, so the higher-level mechanisms (Section~\ref{sec:solver-bestiary}) that introduce phantom hypotheses do not have to reinvent the bookkeeping each time. Third, montage gives the solver a robust way to keep multiple candidate witnesses alive across context changes, which is essential to the cost-budgeted search style of the Miracle runtime (Section~\ref{sec:solver-miracle}).

%% file: appendix/solver.tex
\section{Solver Internals}
\label{sec:solver}

This appendix gives a concrete description of the solver: the rule databases, the rule format, the Miracle runtime that schedules everything under a cost budget, the well-definedness machinery that hangs off of it, and the bestiary of named tactics it dispatches.

\subsection{Rule databases}
\label{sec:solver-rules}

Auto rules are organised into two databases by their direction of application. \emph{Backward} rules live in the \emph{Tekken Tachikaze} database: when a goal $q$ comes up, the solver searches for a rule whose conclusion unifies with $q$ and recurses on the rule's assumptions. \emph{Forward} rules live in the \emph{Ashisogi Jizō} database: when a fresh hypothesis $h$ is derived, the solver searches for a rule whose assumption matches $h$ and adds the rule's conclusion as a new hypothesis. Each database is indexed by a discrimination tree over the rule's pattern shape, so lookup is sub-linear in the number of rules.

\subsection{Rule format}
\label{sec:solver-format}

Rules are written as LaTeX propositions in spec files (one file per rule database under the project's \texttt{specs/} tree) using the same surface syntax as user input:
\begin{quote}\small
\begin{verbatim}
\begin{proposition}
    [rule-name]
    \label{prop:rule_name}
    Let $x, y \in \mathbb{R}$.
    Assume <pattern>.
    Then <conclusion>.
\end{proposition}
\end{verbatim}
\end{quote}
A rule is loaded into the appropriate database after the project's specs are read, and is available to the solver on every subsequent proof obligation. Adding a new rule does not require touching the elaborator source.

\subsection{The Miracle runtime}
\label{sec:solver-miracle}

The solver runs inside a runtime we call \emph{Miracle}, which evaluates a non-deterministic computation under a global cost budget. Each candidate continuation (e.g.\ a different rule fired against the same goal) is enqueued with a fee; the runtime explores in priority order and aborts a branch as soon as its accumulated cost exceeds the budget. Hierarchical decomposition is natural: a sub-obligation gets its own sub-budget. The budget bounds end-to-end work and rules out non-termination, while the per-step fee scheme lets cheap and reliable rules fire first. This cost-budgeted, priority-scheduled search is similar in spirit to the cost system in ProofGrader~\cite{proofgrader}, which manages multiple sub-solvers with per-solver cost budgets and dynamic priority scheduling over hierarchical proof obligations. The name is not coincidence: the connotation is a deliberate, costly intervention --- something the runtime reaches for only when its narrower tactics have not closed the goal.

\subsection{Diagnostic on failure}
\label{sec:solver-diagnostic}

When the budget is exhausted without closing a goal, Miracle reports the goal that was being attempted together with the source location it traces back to. This is the diagnostic an LLM sees in the autoformalizer loop. (A current limitation is that the diagnostic does not yet report which rule almost fired or how close the best candidate came; see Section~\ref{sec:limitations}.)

\subsection{Well-definedness obligations}
\label{sec:solver-wd}

Every user-written expression might carry implicit hypothesis arguments --- proof obligations the elaborator must discharge before the expression is accepted. The elaborator carries a per-operator description of what each partial-function application demands: field division demands that the divisor be non-zero; integer powers of a possibly-negative base demand either that the base be non-zero or that the exponent be non-negative; the real logarithm demands its argument be positive; the inverse of a function-like object demands bijectivity; and the various $a^b$ regimes (real-to-real, real-to-complex, non-negative-base-to-positive-real-exponent, real-to-integer, real-to-natural) each carry the appropriate domain conditions. Function signatures themselves remain plain (division is $\mathbb{R} \to \mathbb{R} \to \mathbb{R}$, as in Lean / Rocq / Mathlib); the obligation set lives in the elaborator, not in the type. Each emitted obligation is added to the current proof state and discharged by the same Miracle runtime that closes ``Then \dots'' gaps; an application whose obligations cannot be closed is rejected with a diagnostic at the offending sub-expression.

\subsection{The normalisation system}
\label{sec:solver-normalization}

Visored carries a unified normalisation pass on expressions that turns every accepted (well-defined) expression into a canonical form. The pass is stronger than what a ring normaliser (\texttt{ring\_nf}) or field normaliser (\texttt{field\_nf}) can do safely without help, because the elaborator has already discharged the well-definedness obligations of every sub-expression: the normaliser is free to carry out deeper reductions that would otherwise be unsound. For instance, once $x \neq 0$ is in the hypothesis state, $x/x$ normalises to $1$ unconditionally; comparable reductions fire across multiplication by an inverse, exponentiation by zero or one, division by literals, $\log$ of a power, and so on. The implementation is intricate --- each operator class carries its own normalisation rules, and the system runs them to a fixed point --- and we do not document the full rule set in this paper.

The normaliser is what makes the rule databases tractable. Without canonical forms, a discrimination tree over rules would have to anticipate every algebraically-equivalent shape of the same expression, and the rule databases would grow combinatorially. With canonical forms, the discrimination tree only needs to match the canonical shape. Normalisation is the system's primary way of making algebraic reasoning cheap.

\paragraph{Cost on the transpilation side.} The same normalisation pass is the largest single source of verbosity in emitted Lean (Appendix~\ref{sec:transpilation}). Normalisation is cheap on the Visored side --- a single fixed-point pass per expression --- but Lean has no comparable normaliser available to call back into (\texttt{ring\_nf} and \texttt{field\_nf} cannot use well-definedness facts to discharge their side conditions), so when an accepted proof is transpiled, the emitter has to spell out the normalisation history one primitive rewrite at a time. The emitted file ends up resembling an interpreter trace of the Visored normaliser running on the input.

\subsection{Asymmetric context propagation across boolean connectives}
\label{sec:solver-asymmetric}

When the elaborator processes $a \land b$, the hypothesis stack is updated with $a$ \emph{before} $b$ is processed, so $b$'s well-definedness obligations are checked under the assumption $a$. Concretely, the conjunction-folding ambush adds the left conjunct to the spotlight before processing the right one, so a fact derivable from $a$ --- say $x \neq 0$ from $1 < x$ --- is available when $b$'s well-definedness is checked. The reverse direction does not hold: writing $a \land b$ does not let one assume $b$ when checking $a$. The same asymmetric flow holds for $a \to b$ (assume $a$ when checking $b$'s well-definedness) and for $\textsf{if } a \textsf{ then } b \textsf{ else } c$ (check $b$ under $a$, $c$ under $\neg a$). This matches the rule PVS adopted in the early 1990s for its TCCs~\cite{pvs-types-doc}, arrived at here independently from a CNL surface.

\subsection{A bestiary of solver mechanisms}
\label{sec:solver-bestiary}

Internal solver mechanisms in Visored carry evocative rather than descriptive names. Every mechanism has its own module, and the module's docstring explains what the code actually does; the names recurring below act as a navigational handle in the source tree. The mechanisms below \emph{call each other recursively} --- a tactic on a goal will invoke other tactics on its sub-goals, and an ambush on a fresh hypothesis can trigger further ambushes --- with the Miracle runtime pruning unproductive branches under its global cost budget. The overall solver behaviour emerges from the interaction of many such mechanisms rather than from any single one.

\paragraph{Tekken Tachikaze 鐵拳断風 --- lightweight backward sweep.}
\emph{Iron-Fist Wind-Cleaver.} A large database of cheap backward-reasoning rules, each firing fast on a goal. Collectively they cover most common deductions; the runtime tries them first because their cost is negligible. One sweep, many small things felled.

\paragraph{Mashiro Kikku マシロキック --- non-recursive trivial close.}
\emph{White Kick.} Like Tekken Tachikaze, but its rule database fires \emph{trivially} and without recursion: a single decisive pass that closes whatever it can immediately, then stops.

\paragraph{Ashisogi Jiz\=o 疋殺地蔵 and Konjiki Ashisogi Jiz\=o 金色疋殺地蔵 --- forward propagation.}
\emph{Foot-Killer Jiz\=o; Golden Foot-Killer Jiz\=o.} Forward propagation on a freshly derived fact: pattern-match on its shape (logical folds, numeric or set chains, applications, quantified expressions) and add every immediate consequence to the hypothesis stack. Once it fires, the consequences spread outward and the affected proposition cannot escape them. The golden variant specialises in chaining-separated lists with exactly two elements.

\paragraph{Hy\=orinmaru 氷輪丸 --- algebra.}
\emph{Ice-Ring Pinwheel.} The algebra solver. It freezes continuous algebraic structure into discrete combinatorial witnesses: integer-interval saturation (cutsat1dx), modular implications, GCD/LCM via prime-factor decomposition, and rationality / non-rationality assertions for radicals.

\paragraph{Ry\=ujin Jakka 流刃若火 --- analysis.}
\emph{Flowing-Blade Young-Flame.} The analysis solver, paired with Hy\=orinmaru: where the ice freezes algebra into discrete combinatorial witnesses, the flame fuses scattered analytic constraints into single sweeping ones. Currently it merges scattered interval constraints into unified intervals, both for continuous real intervals and discrete integer intervals.

\paragraph{Sode no Shirayuki 袖白雪 and Zabimaru 蛇尾丸 --- qualified hypothesis instantiation.}
\emph{Sleeve of White Snow; Snake-Tail Pinwheel.} A family that discharges the goal by drawing on \emph{qualified hypotheses already in the proof context} and instantiating them. Sode no Shirayuki operates at the general expression level, pulling from the \emph{rukia inventory} of explicit and implicit context-rules --- the visible patterns drawn out of the sleeve, and the hidden ones woven beneath. Zabimaru is the chain-specialised counterpart: pulling from the \emph{renji inventory} indexed by the chain's left-hand side, then bridging the remaining gap via the Kurapika link-prover --- a segmented whip-blade each section of which locks onto one link.

\paragraph{Zangetsu 斬月 and Tensa Zangetsu 天鎖斬月 --- congruence reduction.}
\emph{Slaying Moon; Heaven-Chain Slaying Moon.} Equality reduction in the spirit of Lean's \texttt{congr} / \texttt{gcongr}. Zangetsu cuts both sides through a common structure: $\ln x = \ln y \leadsto x = y$, $|x|=|y| \leadsto x = y \lor x = -y$. Tensa Zangetsu is the asymmetric, sharper variant --- it cuts only one side ($\log x = y \leadsto x = \exp y$) and can fire only once per goal.

\paragraph{Senbonzakura 千本桜 and Senbonzakura Kageyoshi 千本桜景厳 --- heavy library search.}
\emph{Thousand Cherry Blossoms; Thousand Cherry Blossoms Vista.} A thorough sweep through the library: Senbonzakura tries each rule against the current target and keeps a match only when the result is syntactically equivalent. The more powerful variant Kageyoshi extends the sweep further, accepting matches via an additional implicational transformation rather than strict equivalence --- a sprawling field of petals where the smaller variant was already a dense one.

\paragraph{Ky\=oka Suigetsu 鏡花水月 --- ephemeral conditional hypothesis.}
\emph{Mirror-Flower Water-Moon.} Proves $\forall x.\,P(x)$, $p \to q$, and the internal imb form by \emph{introducing a phantom hypothesis} --- the bound variable, or the antecedent --- that the system temporarily believes in while it proves the body. Once the body is proved, \emph{ascension} (Appendix~\ref{sec:kernel-ascension}) is what lifts the result past the phantom hypothesis to close the unconditional outer goal; the hypothesis dissolves in the act of ascending. Ascension is critical to this mechanism, which is fitting --- ascent is, in another setting, what this name's wielder is best known for pursuing.

\paragraph{H\=ozukimaru 鬼灯丸 and Ry\=umon H\=ozukimaru 龍紋鬼灯丸 --- two chains into one.}
\emph{Demon-Lantern Pinwheel; Dragon-Crest Demon-Lantern.} ``Obtain one chain from two chains'': merges two chaining-separated lists into a single chain by dispatching on the separator (numeric, propositional, or set) and applying a joining rule. Ry\=umon H\=ozukimaru, the more powerful variant, runs the same merge with \texttt{try\_apply\_new} instead of \texttt{try\_apply\_new\_trivially}, allowing the join to fire on a non-trivial rule application.

\paragraph{Kazeshini 風死 and Kazeshini Fushi no K\=oj\=o 風死・不死の工場 --- weaken existing hypothesis.}
\emph{Wind-Death; Wind-Death Immortal Factory.} Discharges the goal by \emph{weakening} an existing context hypothesis at wind-like speed: $x > 5$ weakens to $x \ge 5$, $x < y$ weakens to $x \ne y$, and so on, with sub-cases for propositions, numerics, sets, and modular equivalences. The Immortal Factory variant restricts the weakening search to \emph{primary} hypotheses --- the first-principles materials of the context --- rather than the full assumption stack.

\paragraph{Nozarashi 野晒 --- split the goal.}
\emph{Sun-Bleached.} Cuts the goal into smaller goals when its shape allows: a conjunction splits into both conjuncts, a disjunction into a side-choice, an implication or iff splits via logic-arrow rules, and a roster subset goal splits into per-element memberships. A weather-worn cleaver that hews the goal's grain in two.

\paragraph{Sakanade 逆撫 and Sakashima Yokoshima Happ\=ofusagari 逆様邪八方塞がり --- stroke against the grain.}
\emph{Reverse-Stroke; Backwards-and-Crooked Eight-Direction-Sealed.} Both reverse rather than approach head-on. \textbf{Sakanade} is the ambush form: when a fresh hypothesis arrives, generate its reversed/contrapositive form as a new hypothesis. \textbf{Sakashima Yokoshima Happ\=ofusagari} is the tactic form: reverse the current goal into an equivalent reversed shape, prove the reversed shape, and wrap the result in a Sakanade hypothesis construction. The longest name in the source tree evokes the complete inversion --- the goal turned about, every direction sealed off.

\paragraph{Benihime 紅姫 --- extensible simp-style normalisation.}
\emph{Crimson Princess.} On a fresh hypothesis, Benihime applies a simp-style rule database to rewrite it into a canonical normal form before it enters the hypothesis stack. The matching tactic Benihime serves as the gateway through which the simp pass is invoked --- a multifaceted entry point, extensible across many small refining disciplines.

\paragraph{Kage Bunshin no Jutsu 影分身の術 --- unchecked-rule escape hatch.}
\emph{Shadow-Clone Technique.} A hypothesis-side ambush that applies a rule Visored does not internally re-check but trusts a downstream macro to re-verify, decohering a superposed proposition into a single concrete component. The genuine consequence is kept; the clones are discarded.

\paragraph{The Kurapika クラピカ family --- graph algorithms over chains.}
Named after an analytical chain-master. \textbf{Kurapika Chain} uses graph algorithms to congregate sub-chains and prove a chain-separated goal end-to-end. \textbf{Kurapika Link} proves individual links by saturating the hypothesis stash and bridging gaps with transitivity. \textbf{Kurapika Substitute} reforges expressions by substituting along active arrows in the chain graph. The \textbf{Kurapika} ambush splits a non-trivial chain into left and right halves and stashes both for the others to assemble.

%% file: appendix/transpilation.tex
\section{Transpilation}
\label{sec:transpilation}

This appendix describes how an accepted Visored proof is re-expressed as Lean source: the intermediate representation, the two emission schemes, the coercion bookkeeping that drives the file size, the discharge-primitive heuristic, and the supporting Lean library. The high-level picture is that transpilation is a separate stage from elaboration and proof search, and the design choices here are independent of Visored's correctness as a checker.

\subsection{UVL MIR: a target-neutral IR}
\label{sec:transpilation-uvl}

Visored does not emit Lean directly. An accepted proof is first lowered into UVL MIR, a target-neutral intermediate from which a target-specific formatter emits source text. The UVL representation carries each derivation step as an abstract operation together with the types and witnesses it needs; the Lean formatter realises each step as a Lean term or tactic call. The benefit of this split is that adding a new prover target (Rocq, Isabelle, \dots) is a new formatter on top of UVL rather than a rewrite of elaboration. In principle the difficulty of transpiling to Rocq or Isabelle is comparable to transpiling to Lean: the same coercion bookkeeping, the same discharge-primitive heuristic, and the same supporting-library cost would reappear on each new target. Only the Lean target has been implemented so far. The cost of the split is that the UVL layer carries information that no single target needs in its entirety, so the formatter has to make small choices at each step about what to materialise and what to leave implicit.

\subsection{Sparse versus dense emission}
\label{sec:transpilation-modes}

There are two emission schemes with opposite trade-offs.

The \emph{sparse} scheme, which is the current default, realises each Visored derivation step as its own \texttt{private lemma}, with the step's inputs and outputs spelled out explicitly. Every step is independently re-checkable by Lean and a diagnostic can point at the exact failing lemma; the cost is the file-size figures discussed below.

The \emph{dense} scheme bundles a contiguous run of steps inside a single \texttt{by \dots} tactic block. The emitted source is much more compact and Lean compiles it faster, but the per-step diagnostic channel collapses to whatever Lean reports for the block as a whole.

We currently default to sparse; the choice is a debuggability-versus-size trade-off rather than a permanent design commitment.

\subsection{Replaying the normalisation history}
\label{sec:transpilation-normalization}

The largest single contributor to emitted size is the cost of replaying Visored's expression normalisation system (Section~\ref{sec:solver-normalization}). On the Visored side normalisation is cheap: a single fixed-point pass reduces every accepted expression to a canonical form, and the rule databases match against canonical shapes. Lean has no comparable normaliser available to call back into --- \texttt{ring\_nf} and \texttt{field\_nf} are weaker because they cannot use the well-definedness facts the Visored elaborator has already discharged --- so when the emitter has to express ``Visored normalised these two expressions to the same canonical form'', it has to spell out the normalisation history one primitive rewrite at a time. Each rewrite becomes its own named Lean lemma; the emitted file is essentially an interpreter trace of the Visored normaliser running on the input. This single phenomenon is the root cause of the line-count ratios reported below. 
Section~\ref{sec:transpilation-example} below shows a small slice of an actual emitted file.\footnote{Representative emitted-Lean excerpts---the first chunks of one accepted proof per miniF2F subject, each
  annotated with the full emission size---are available at \url{https://github.com/xiyuzhai-husky-lang/visored}.}

\subsection{Coercion bookkeeping}
\label{sec:transpilation-coercion}

The secondary contributor is coercion bookkeeping between number systems. Visored expressions are typed bottom-up at the least type the construction naturally lives in (Appendix~\ref{sec:semantics}); Lean elaborates top-down and inserts coercions at unification-driven positions. The two views are reconciled in two places.

On the Lean side, the \texttt{pull\_cast\_rewrite} pass rewrites ML-style Lean expressions into the bottom-up canonical form before any Visored-derived lemma is applied.

On the emission side, every Visored derivation kind that crosses numeric types (\(\mathbb{N} \leftrightarrow \mathbb{Z} \leftrightarrow \mathbb{Q} \leftrightarrow \mathbb{R} \leftrightarrow \mathbb{C}\)) is realised as a generic lemma that takes explicit coercion-witness arguments at the call site --- witnesses like \texttt{nat\_int\_int\_coercion\_triangle} or \texttt{num\_eq\_nat\_to\_int\_coercion}, each pinning down one piece of the cast tower. This is correct and uniform across number-system combinations, but it compounds with the per-rewrite emission of the normaliser trace: every replayed normalisation step that crosses a numeric-type boundary additionally carries its own packet of coercion witnesses.

\subsection{Discharge primitives, stock and custom}
\label{sec:transpilation-discharge}

For sub-goals that are not pure-symbolic --- numeric inequalities, ring identities, polynomial equations, modular and divisibility constraints, small finite-case verifications --- the emitter routes to a discharge primitive. Some primitives are stock Lean / Mathlib tactics (\texttt{norm\_num}, \texttt{ring}, \texttt{decide}, \texttt{native\_decide}, \texttt{simp}). For shapes that the stock primitives do not handle well, Visored introduces its own Lean tactics: a custom \texttt{cutsat1dx} tactic for integer linear arithmetic with explicit cut certificates, a polynomial-equation tactic for Hy\=orinmaru-side derivations, and a strengthened \texttt{norm\_num\_extra} for numeric comparisons the standard \texttt{norm\_num} times out on. Choosing among these is a heuristic on the emitter side: pick the lightest primitive observed to close this shape of goal. Wrong choices typically produce a Lean file that still type-checks but takes an order of magnitude longer for \texttt{lake build} to verify, or one that fails with a diagnostic far from the underlying mistake. Improving the heuristic is one of the smaller but high-impact items on the transpilation roadmap.

\subsection{The Visored Lean library}
\label{sec:transpilation-library}

Emitted Lean source calls into a supporting Lean library of pre-proved lemmas, named tactics, and macros. The current snapshot contains roughly $3{,}600$ theorems and lemmas across about $280$ files and $46{,}000$ lines, plus roughly $170$ custom tactic / syntax / macro declarations. The bulk breaks down approximately as:

\begin{itemize}
\item $\sim 1{,}030$ derivation-side lemmas --- the per-rewrite primitives the normaliser-trace emitter calls into;
\item $\sim 1{,}130$ tactic-side lemmas, including the custom \texttt{cutsat1dx}, polynomial-equation, and \texttt{norm\_num\_extra} machinery;
\item $\sim 660$ coercion lemmas --- the witness tower across $\mathbb{N} \leftrightarrow \mathbb{Z} \leftrightarrow \mathbb{Q} \leftrightarrow \mathbb{R} \leftrightarrow \mathbb{C}$;
\item $\sim 370$ ambush-side lemmas;
\item the remainder split among hypothesis manipulation, foundations, and prelude.
\end{itemize}

A large fraction of these proofs were obtained by a deliberately simple pipeline: the Visored side emits each needed Lean theorem with a \texttt{sorry} body, and Claude / Codex then fills in the \texttt{sorry}s with actual proofs that \texttt{lake build} accepts. There is no clever search loop on top --- the LLM is given the theorem statement and asked to prove it, one lemma at a time. This is one of the load-bearing payoffs of building the system at all. Each Visored CNL construct stands on top of many Lean-side proofs, but those proofs are written once and shared across every accepted CNL example that uses the construct. From the user's perspective, the Visored surface is strictly higher-level than the corresponding Lean surface --- the cost paid once inside the library is what makes the upstream CNL succinct, even though every accepted proof still bottoms out at the Lean kernel.

Every new Visored derivation kind requires both an emitter case and a corresponding Lean-side proof in this library, which is the main reason expanding Visored's surface vocabulary remains a two-sided engineering cost.

\subsection{Verbosity in practice}
\label{sec:transpilation-verbosity}

The sparse scheme produces large files. On the miniF2F-valid set (244 problems, 32 emitting non-trivial Lean in the current snapshot), per-problem line-count ratios are heavy-tailed: median $\approx 250\times$, mean $\approx 1{,}170\times$, ranging from $\sim 5\times$ on the smallest accepted proofs up to $\sim 24{,}800\times$ on the largest (a $6$-line CNL input expanded to a $148{,}889$-line Lean file containing $37{,}416$ named lemmas). The bulk ratio across all measured emissions is $\sim 1{,}100\times$; the mean emitted file contains roughly $2{,}000$ named lemmas, and several typical files sit near $700$ named lemmas each. This is the cost of the sparse scheme's per-step explicitness, and it is the headline reason a denser emission scheme is in progress.

\subsection{Example slices of transpilation}
\label{sec:transpilation-example}

To make the per-rewrite shape concrete, the snippets below are drawn directly from emitted Lean files on miniF2F-valid. We show four flavours of lemma that recur throughout every emission: a basic derivation step with the full coercion-witness packet, a commutative-ring rewrite, a numeric discharge through Visored's custom \texttt{norm\_num\_extra}, and a library-rule application that surfaces a named Visored tactic (Senbonzakura, in this case) directly in the emitted Lean.

\paragraph{Basic derivation with coercion witnesses.} From a problem whose CNL goal is the one-liner $3! \cdot (2^{3}+\sqrt{9})/2 = 33$ (Visored accepts it and emits $318$ named lemmas):

\begin{quote}\scriptsize
\begin{lstlisting}[language=Lean4]
/-- `9 + -0 = 9` by `derivation.term.add_eq`. -/
private lemma h1d5
  : (((9:ℕ) : ℤ) + (-((0:ℕ) : ℤ) : ℤ) : ℤ) = (9:ℕ)
  := Visored.Library.Derivation.Term.add_eq ℕ ℤ ℤ ℕ h1d2 h1d3 h1d4
       num_eq_nat_to_int_coercion nat_nat_int_coercion_triangle
       num_eq_identity_coercion nat_int_int_coercion_triangle
       num_eq_nat_to_int_coercion nat_nat_int_coercion_triangle
       comm_ring_add_nat_to_int_coercion nat_nat_int_coercion_triangle
       nat_nat_int_coercion_triangle
\end{lstlisting}
\end{quote}

\noindent The conclusion is the trivial fact $9 + (-0) = 9$, but the body carries a packet of nine coercion witnesses pinning down every leg of the $\mathbb{N} \to \mathbb{Z}$ cast tower.

\paragraph{Commutative-ring rewrite.} A multiplication-side reforge in the same family, applied to a real-valued expression:

\begin{quote}\scriptsize
\begin{lstlisting}[language=Lean4]
private lemma h3d5 (b : ℝ)
  : ((((4:ℕ) : ℝ) : ℝ) * (b ^ (2:ℕ) : ℝ) : ℝ) = (((4:ℕ) : ℝ) * (b ^ (2:ℕ) : ℝ) : ℝ)
  := Visored.Library.Derivation.ExprReforge.comm_ring_mul_reforge ℝ ℝ ℝ
       h3d3 (h3d4 b)
       num_eq_identity_coercion num_eq_identity_coercion
       nat_real_real_coercion_triangle real_real_real_coercion_triangle
       comm_ring_mul_identity_coercion
       nat_real_real_coercion_triangle real_real_real_coercion_triangle
\end{lstlisting}
\end{quote}

\paragraph{Numeric discharge through Visored's \texttt{norm\_num\_extra}.} When a sub-goal reduces to a comparison the stock \texttt{norm\_num} cannot close, the emitter falls back to Visored's custom \texttt{norm\_num\_extra}:

\begin{quote}\scriptsize
\begin{lstlisting}[language=Lean4]
/-- `9 ≥ 0 ↔ True` by `derivation.term.trivially_true_num_comparison`. -/
private lemma h1d8 : (9:ℕ) ≥ (0:ℕ) ↔ True :=
  Visored.Library.Derivation.Term.trivially_true_num_comparison h1d7
    (by norm_num_extra)
\end{lstlisting}
\end{quote}

\paragraph{A named Visored tactic surfacing in emitted Lean.} The Visored-side dispatch machinery shows up in the emitted code as library calls into the corresponding tactic namespace. For example, the Senbonzakura library rule \emph{squared\_pos\_if\_base\_nonzero} is invoked directly inside an \texttt{apply\_exact\_rule}:

\begin{quote}\scriptsize
\begin{lstlisting}[language=Lean4]
private lemma h2 (b : ℝ) (h0 : b > 0) : (b ^ (2:ℕ) : ℝ) > ((0:ℕ) : ℝ) :=
  Visored.Library.Derivation.ApplyRule.apply_exact_rule
    (Visored.Library.Tactics.Senbonzakura.«prop:squared_pos_if_base_nonzero» b h0)
    (h2d0 b)
\end{lstlisting}
\end{quote}

\noindent A Tekken-Tachikaze-side rule appears in the same shape:

\begin{quote}\scriptsize
\begin{lstlisting}[language=Lean4]
private lemma h_ne : p ≠ ((0:ℕ) : ℤ) :=
  Visored.Library.Tactics.TekkenTachikaze.ne_if_gt ℤ h4
\end{lstlisting}
\end{quote}

These four shapes --- elementary derivation, commutative-ring rewrite, custom numeric discharge, and named-tactic library application --- account for almost every line of a typical emitted file. The shape is straightforward; what makes the line counts heavy is that each rewrite step in the normaliser's history produces one such lemma, and a goal of even modest complexity replays many hundreds of them.

\paragraph{State of the transpilation system.} The current transpilation layer is not yet clean. The sparse emission scheme, the per-operator coercion-witness packets, the discharge-primitive heuristic, the supporting Lean library, and the interaction between them have all evolved incrementally; they are not yet organised in a way we would call clean. A simplification pass is in progress, with the primary targets being denser emission (fewer named lemmas per Visored step), a smaller and more uniform coercion-witness surface, and a re-organisation of the supporting library along the same lines as the elaborator-side classification. We expect the published numbers to improve substantially over the next iteration of the system, and the design described in this appendix to be the last verbose generation rather than the steady state.

%% file: appendix/skill.tex
\section{The Visored Agent Skill}
\label{sec:skill}

This appendix reproduces, verbatim, the skill given to the LLM coding agent in the experiment of Section~\ref{sec:experiments}: the top-level \texttt{SKILLS.md} in full, followed by a representative subfile from each of the skill's three documentation categories (tactics, syntax, workarounds).

\paragraph{SKILLS.md.}
\begin{lstlisting}[basicstyle=\ttfamily\tiny,breaklines=true,columns=fullflexible,extendedchars=true]
---
name: visored-prover
description: Prove math theorems using the Visored controlled natural language theorem prover. Use when working with .tex files containing \begin{example}...\end{example} blocks with math proofs.
---

# Visored Theorem Prover

Visored is a controlled natural language theorem prover. You extend problem prefixes to form complete proofs.

## Workflow

1. **Read the problem** from a `.tex` file (the prefix inside `\begin{example}...\end{example}`)
2. **Think through the proof** in natural language first
3. **Extend the prefix** with proof statements
4. **Verify** by running: `visored-core-cli <file_path> --specs-dir <specs_dir>`
5. **If error**, parse the error message and fix the text
6. **Repeat** until SUCCESS

## Harder Problems

For harder problems:

1. **Solve in natural language first** - Write out the complete proof in plain English/math before attempting CNL
2. **Convert to CNL** - Only after you have a working natural language proof, translate it step by step
3. **If you can't solve it in natural language** - That's YOUR problem, not visored's. Don't blame the prover for your inability to find the proof.

This ensures you distinguish between:
- **Your failure**: Can't find the proof strategy
- **Visored limitation**: Found the proof but visored can't verify a specific step

## Verification Command

**IMPORTANT: Always use `cargo run`, NOT the pre-built binary!**

```bash
# ALWAYS use this form:
cargo run --bin visored-core-cli -- <path_to_tex_file> --specs-dir /home/xiyuzhai/repos/husky2/specs --output-style llm

# Syntax check only (faster):
cargo run --bin visored-core-cli -- <path_to_tex_file> --specs-dir /home/xiyuzhai/repos/husky2/specs --stages syntax_only
```

**NEVER use the pre-built binary directly!**

- SUCCESS → proof is valid
- FAILED → fix the error shown and retry

## Error Format

Errors mark the exact location with `【】` brackets:
```
ERROR: missing embedded math right delimiter
Line 3: We have 【$】\left(...\right) = 36.
```

---

# Language Reference

## Formatting

One sentence per line, separated by blank lines:

```latex
Let $x\in\mathbb{R}$.

Assume $x > 0$.

The goal is to prove $x^2 > 0$.

We have $x^2 > 0$.
```

## Basic Constructs

| Construct | Example | Description |
|-----------|---------|-------------|
| Variable introduction | `Let $x \in \mathbb{R}$.` | Introduce a variable |
| Variable definition | `Let $x = 5$.` | Define a variable with value |
| Assumption | `Assume $x > 0$.` | Add an assumption |
| Goal statement | `The goal is to prove $x^2 \geq 0$.` | State what to prove |
| Assertion | `We have $x^2 \geq 0$.` | Assert a fact (auto-proved) |
| Conclusion | `Then $x^2 \geq 0$.` | Conclude from previous steps |
| Divisibility | `Then $a \mid b$.` | Assert a divides b |
| Divisor count | `$\numberOfDivisors{n}$` | Number of divisors of n (renders as τ(n)) |
| Informal comment | `\informal{explanation}` | Add readable comment (ignored by prover) |

## Universal Instantiation

When you have `∀x, p(x)` and want to prove `q(x_0)` where it follows from `p(x_0)`, **explicitly state `p(x_0)` first!**

```latex
% Given: ∀n∈ℕ, a(n+2) = a(n)·a(n+1)/(2a(n)-a(n+1))
% Want to use it for n=1

% WRONG - trying to use consequence directly:
We have $a(3) = \frac{3}{11}$.  % Fails! Visored doesn't auto-instantiate

% RIGHT - explicitly instantiate first:
We have $a(3) = \frac{a(1)\cdot a(2)}{2a(1)-a(2)}$.  % Instantiate ∀ with n=1
We have $a(3) = \frac{3}{11}$.  % Now compute the value
```

**Don't be lazy!** Always explicitly write out the instantiation before using its consequences.

## Bounded Variable Introduction

To prove a universal statement `∀k, P(k) → Q(k)`, use the "Let such that" pattern:

```latex
Let $k \in \mathbb{Z}$ such that $1 \le k \le n$.

<prove property for k>

Then $\forall k \in \mathbb{Z},\, 1 \le k \le n \implies <property>$.
```

Key points:
- `Let $k \in S$ such that <condition>` introduces k with the bound attached
- Prove the property for that generic k
- `Then $\forall k \in S,\, <condition> \implies <property>$` generalizes

**IMPORTANT:** This pattern works inside case branches where `Let`+`Assume` would fail.

## Chaining

Use chaining `A = B ≤ C` to improve readability and help Visored reasoning:

```latex
% Instead of separate steps:
We have $\sum_{k=1}^{10} f(k) = \sum_{k=1}^{5} f(k) + \sum_{k=6}^{10} f(k)$.
Then $\sum_{k=1}^{10} f(k) = 10 + \sum_{k=6}^{10} f(k)$.
Then $\sum_{k=1}^{10} f(k) \leq 10 + 20$.

% Use chaining:
Then $\sum_{k=1}^{10} f(k) = \sum_{k=1}^{5} f(k) + \sum_{k=6}^{10} f(k) \leq 10 + 20$.
```

Chaining works with `=`, `<`, `≤`, `>`, `≥` in any combination.

**For inequalities with variable bounds:** Break into steps - one for substitution, one for simplification:
```latex
% Instead of: Then $(n-1) \cdot n \cdot (n+1) \ge 990$.
% Use chaining with explicit substitution:
Then $(n-1) \cdot n \cdot (n+1) \ge (10-1) \cdot 10 \cdot (10+1) = 990$.
```

**For complex sums/products:** When substitution fails on nontrivial sum expressions, use chaining to combine the steps into one statement.

## Disjunction Elimination

When you have `P ∨ Q` and want to derive a result, use case analysis:

```latex
Assume $a = 2 \lor b = 2$.
Assume $a \neq 2$.
The goal is to prove $b = 2$.
We prove $b = 2$ by working through different cases :
\begin{itemize}
    \item Case $a = 2$. Contradiction.
    \item Case $b = 2$. Then $b = 2$.
\end{itemize}
```

Key: Use `Contradiction.` when a case is impossible given other assumptions.

**For factored equations:** When you have `(A)(B) = 0`, use case analysis on `A = 0 ∨ B = 0`:

```latex
Then $(2a + 1)(a - 1) = 0$.
Then $2a + 1 = 0 \lor a - 1 = 0$.
We prove $a = 1$ by working through different cases :
\begin{itemize}
    \item Case $2a + 1 = 0$. Then $a = -\frac{1}{2}$. Then $a < 0$. We have $a > 0$. Contradiction.
    \item Case $a - 1 = 0$. Then $a = 1$.
\end{itemize}
```

Use contradiction to eliminate impossible cases based on other constraints (e.g., `a > 0`).

**IMPORTANT:** Always use `Contradiction.` instead of `Then $\mathsf{False}$.` The contradiction must be explicit in the proof - write out the conflicting statements (e.g., `We have $n \ge 55$. We have $n < 55$.`) before `Contradiction.`

## Induction

Use `We prove $P(n)$ by induction on $n$ :` with `\begin{itemize}` cases. See `docs/tactics/induction.md`.

**IMPORTANT: Induction must start from 0.** If you need to prove something for `n ≥ k` where `k > 0`, use a substitution:

```latex
% If you need to prove P(n) for n ≥ 1:
Let $m \in \mathbb{N}$.
We prove $P(m+1)$ by induction on $m$ :  % Now m starts from 0
\begin{itemize}
    \item Case $m = 0$. ... prove P(1) ...
    \item Case $m \ge 0$. Assume $P(m+1)$. ... prove P((m+1)+1) ...
\end{itemize}
```

This transforms `n = m + 1` so that when `m = 0, 1, 2, ...`, we get `n = 1, 2, 3, ...`

## Claim `∈ ℕ` Before Using Quotients in Auto Rules

**IMPORTANT:** When using a quotient like `n/k` in auto rules (e.g., `dvd-gcd-if-dvd-both`), you must first establish that `n/k ∈ ℕ`. Otherwise the auto rule matching will fail silently.

```latex
% WRONG - auto rule can't match n/5 without knowing it's in ℕ:
Assume $5 \mid n$.
Assume $n / 5 \mid 3628800$.
Assume $n / 5 \mid n$.
Then $n / 5 \mid \gcd(3628800, n)$.  % FAILS

% RIGHT - claim n/5 ∈ ℕ first:
Assume $5 \mid n$.
Then $n / 5 \in \mathbb{N}$.
Then $n / 5 \mid 3628800$.
Then $n / 5 \mid n$.
Then $n / 5 \mid \gcd(3628800, n)$.  % WORKS
```

## Comma Precedence in Set Builder

**IMPORTANT:** `,` has higher precedence than `\mid` (divisibility), but `,\,` has lower precedence.

If you get a `todo!()` panic at `decohere`, it may be because you used `,` instead of `,\,` between propositions in set builder notation.

```latex
% WRONG - comma has higher precedence than \mid, causes parse error:
\{n \in \mathbb{N} \mid 3000 \mid n, n \mid 18144000\}

% RIGHT - use ,\, which has lower precedence:
\{n \in \mathbb{N} \mid 3000 \mid n,\, n \mid 18144000\}
```

## Conjunction in Universal Bodies

**IMPORTANT:** Within the body of `\forall`, use `\land` (not `,\,`) to conjoin conditions. Using `,\,` causes conditional well-definedness to fail — guards won't propagate to later terms.

```latex
% WRONG - guards don't propagate for well-definedness:
\forall n\in\mathbb{N},\, t(n)\neq 0,\, \frac{1}{25t(n)}=1

% RIGHT - \land lets guards propagate:
\forall n\in\mathbb{N},\, t(n)\neq 0 \land \frac{1}{25t(n)}=1
```

The first `,\,` after the quantifier domain (`\forall n\in\mathbb{N},\,`) is fine — it separates the binding from the body. But within the body itself, use `\land`.

## Syntax Rules

**IMPORTANT: `$...$` must enclose a complete syntax tree.** Never break math expressions across boundaries.

**Trigonometric function powers:** e.g., use `{(\sin x)}^n` instead of `\sin^n x`.

```latex
% WRONG - breaks syntax tree:
We have $a = 4 \rightarrow a$ is prime.

% RIGHT - complete expression:
We have $4$ is prime.  % separate statement
```

**Avoid excessive implications.** Write conclusions directly. Use case analysis with `Contradiction.` instead of chaining implications.

## Linear Systems

**Visored does NOT solve linear systems automatically.** Use Gaussian elimination externally (e.g., Python) to find coefficients.

For a system of n equations, find a linear combination that yields the target expression:

```python
from fractions import Fraction
# Solve for coefficients c1, c2, ... such that
# c1*eq1 + c2*eq2 + ... = target_expression
# Use Gaussian elimination with exact Fraction arithmetic
```

Then write the proof showing the linear combination equals the target:

```latex
We have $c_1 \cdot (\text{eq1}) + c_2 \cdot (\text{eq2}) + \ldots = x^2 + y^2$.

We have $c_1 + c_2 + \ldots = 36$.

Then $x^2 + y^2 = 36$.
```

See `solutions/solution75.tex` for a complete example (AIME 1984 P15).

---

# Tactics Reference

See `docs/tactics/` for detailed tactics:

- `case_analysis.md` - Case analysis and contradiction
- `floor.md` - Floor function (Hermite's identity)
- `induction.md` - Induction proofs
- `modular.md` - Modular arithmetic (prefer `\pmod` for derivations, see below)
- `set.md` - Set minimality, forall statements, set equality

## Modular Arithmetic Pattern

**In derivations, prefer `\pmod` (congruence) over `\bmod` (remainder).**

When a problem uses `\bmod` syntax, derive using `\pmod` then convert back:

```latex
% Given: m \bmod 2 = 1
% Want to derive: (m+1) \bmod 2 = 0

% Step 1: Convert to pmod
We have $m \equiv 1 \pmod{2}$.

% Step 2: Derive using pmod (this is what Visored can do)
Then $m+1 \equiv 0 \pmod{2}$.

% Step 3: Convert back to bmod
Then $(m+1) \bmod 2 = 0$.
```

---

# Directory Structure

```
valid/
├── SKILLS.md             # This file
├── progress.json         # Tracks solved/stuck problems
├── docs/
│   ├── tactics/          # Proof tactics (case analysis, sets, etc.)
│   ├── syntax/           # Correct LaTeX syntax rules
│   └── workarounds/      # Current limitations needing manual handling
├── scripts/
│   └── update_progress.py # Progress tracking script
├── problems/             # Problem prefixes
└── solutions/            # Completed proofs
```

## When Stuck

1. Check `docs/tactics/` for proof strategies
2. Check `docs/syntax/` for LaTeX syntax rules
3. Check `docs/workarounds/` for known limitations
4. If new issue, report to maintainer and document

## Reference Examples

- `test-data/visored/elaborator/` - Extensive examples
  - `main/props/` - Basic propositions
  - `minif2f/valid/` - Competition math problems
  - `tactics/` - Various proof tactics
\end{lstlisting}

\paragraph{docs/tactics/set.md.}
\begin{lstlisting}[basicstyle=\ttfamily\tiny,breaklines=true,columns=fullflexible,extendedchars=true]
# Set Tactics

## Set Minimality (Smallest Element)

To prove `$x$ is the smallest element of $S$`:

1. Prove `$x \in S$`
2. Prove `$\forall y \in S,\, x \le y$`

### Example

```latex
Let $S\subseteq\mathbb{Q}$.
Let $x\in\mathbb{R}$.
Assume $x\in S$.
Assume $\forall y\in S,\,x\le y$.
Then $x$ is the smallest element of $S$.
```

## Second Smallest Element

To prove `$x_2$ is the second smallest element of $S$`:

1. Prove `$x_1 \in S$`
2. Prove `$x_1$ is the smallest element of $S$`
3. Prove `$x_2 \in S$`
4. Prove `$\forall y \in S,\, x_1 < y \rightarrow x_2 \le y$`

### Example

```latex
Let $S\subseteq\mathbb{Q}$.
Let $x_1, x_2\in\mathbb{R}$.
Assume $x_1\in S$.
Assume $x_1$ is the smallest element of $S$.
Assume $x_2\in S$.
Assume $\forall y\in S,\,x_1 < y \rightarrow x_2\le y$.
Then $x_2$ is the second smallest element of $S$.
```

Reference: `test-data/visored/elaborator/tactics/tekken_tachikaze/main.tex`

## Proving Forall Statements

To prove `$\forall y \in S,\, P(y)$` (where S must be nonempty):

```latex
Let $y \in S$.

... (proof steps) ...

Then $P(y)$.

Then $\forall y \in S,\, P(y)$.
```

To prove `$\forall y \in S,\, Q(y) \rightarrow P(y)$` (forall with implication):

```latex
Let $y \in S$ such that $Q(y)$.

... (proof steps) ...

Then $P(y)$.

Then $\forall y \in S,\, Q(y) \rightarrow P(y)$.
```

Use `Let $y \in S$ such that $Q(y)$` to introduce an element with an additional condition.

## Set Equality via Predicate Equivalence

To prove `{n ∈ S | P(n)} = {n ∈ S | Q(n)}`, first prove `∀n ∈ S, P(n) ↔ Q(n)`, then conclude set equality.

```latex
% Want: S = {n ∈ ℕ | Q(n)} where S = {n ∈ ℕ | P(n)}
Let $n \in \mathbb{N}$.

... (prove P(n) ↔ Q(n)) ...

Then $\forall n \in \mathbb{N},\, P(n) \iff Q(n)$.

Then $S = \{n \in \mathbb{N} \mid Q(n)\}$.
```

## Set Equality with Roster (Enumerated Set)

To prove $S = \{a, b, ...\}$ or $|S| = n$:

1. **Show each element is in S** (forward direction)
2. **Show all elements of S satisfy a characterization** (backward direction)
3. **Conclude the iff characterization**
4. **Derive set equality and cardinality**

```latex
% Show 1 ∈ S, 2 ∈ S
Then $1 \in S$.
Then $2 \in S$.

% Show ∀x ∈ S, x = 1 ∨ x = 2
Let $x \in S$.
Then ... (derive bounds on x)
We have $x \in \mathbb{N}$.
We have $x > 0$.
We have $x < 3$.
Then $x = 1 \lor x = 2$.

% Conclude
Then $\forall x \in S,\, x = 1 \lor x = 2$.
Then $\forall x \in \mathbb{N},\, x \in S \iff x = 1 \lor x = 2$.
Then $S = \{1, 2\}$.
Then $|S| = 2$.
```
\end{lstlisting}

\paragraph{docs/tactics/induction.md.}
\begin{lstlisting}[basicstyle=\ttfamily\tiny,breaklines=true,columns=fullflexible,extendedchars=true]
# Induction

For proving `∀n∈ℕ, P(n)` by induction:

```latex
We prove $\forall n\in\mathbb{N},\, P(n)$ by induction on $n$ :

\begin{itemize}

\item Case $n=0$.

<proof that P(0) holds>

\item Case $n \ge 0$.

Assume $P(n)$.

The goal is to prove $P(n+1)$.

<proof using inductive hypothesis>

\end{itemize}
```

## Example: Prove 3 | n³+2n

```latex
We prove $3\mid n^3+2n$ by induction on $n$ :

\begin{itemize}

\item Case $n=0$.

We have $3\mid 0$.

\item Case $n \ge 0$.

Assume $3\mid n^3+2n$.

The goal is to prove $3\mid {(n+1)}^3+2(n+1)$.

We have ${(n+1)}^3+2(n+1) = (n^3+2n) + 3(n^2+n+1)$.

We have $3\mid n^3+2n$.

We have $3\mid 3(n^2+n+1)$.

Then $3\mid {(n+1)}^3+2(n+1)$.

\end{itemize}
```

See `test-data/visored/core/lean4-ok/minif2f-maiden-voyage/los_lobos000193.tex` for more examples.
\end{lstlisting}

\paragraph{docs/syntax/exponent-with-parentheses.md.}
\begin{lstlisting}[basicstyle=\ttfamily\tiny,breaklines=true,columns=fullflexible,extendedchars=true]
# Exponent with Parenthesized Base

When raising a parenthesized expression to a power, wrap in braces.

## Error

```
ERROR: right delimiter as base
```

## Wrong

```latex
$(x+3)^2$
$a(x+3)^2$
```

## Correct

```latex
${(x+3)}^2$
$a{(x+3)}^2$
```

## Why

LaTeX parsing requires braces to group the base when it ends with a delimiter like `)`.
\end{lstlisting}

\paragraph{docs/workarounds/quantified-equality-transitivity.md.}
\begin{lstlisting}[basicstyle=\ttfamily\tiny,breaklines=true,columns=fullflexible,extendedchars=true]
# Quantified Equality Transitivity

Visored doesn't auto-chain transitivity for quantified equalities.

## Symptom

```
ERROR: Unable to prove: \(\forall x, A = C\)
```

When you have `∀x, A = B` and `∀x, B = C` established.

## Note

**Non-quantified** equalities work automatically:
- `A = B`, `B = C` → visored derives `A = C`

**Quantified** equalities need explicit steps:
- `∀x, A = B`, `∀x, B = C` → must explicitly state `∀x, A = C`

## Fix

State each transitivity step explicitly:

```latex
We have $\forall x, A = B$.
We have $\forall x, B = C$.
We have $\forall x, A = C$.  % explicit transitivity step
We have $\forall x, C = D$.
We have $\forall x, A = D$.  % another explicit step
```

## Example (Problem 10)

```latex
We have $\forall x \in \mathbb{R},\,3x^2+7x+4 = a{(x+3)}^2 + b(x+3) + c$.
We have $\forall x \in \mathbb{R},\,a{(x+3)}^2 + b(x+3) + c = ax^2 + 6ax + 9a + bx + 3b + c$.
We have $\forall x \in \mathbb{R},\,3x^2+7x+4 = ax^2 + 6ax + 9a + bx + 3b + c$.
```
\end{lstlisting}

%% file: appendix/examples.tex
\section{Additional Worked Examples}
\label{sec:more-examples}

The examples in this appendix are paired Visored CNL / Lean 4 artifacts maintained alongside this paper's source. The CNL on the left is the exact text the Visored elaborator accepts; the Lean on the right is a hand-written Lean 4 proof of the same statement. A \texttt{Makefile} re-runs both checks. All examples are drawn from miniF2F. Within each subject category, examples are ordered roughly easy to hard.

\subsection{Algebra}
\label{sec:ex-algebra}

  \par\addvspace{\medskipamount}%
  \noindent%
  \begin{minipage}[t]{0.47\textwidth}%
  \textbf{Visored CNL input (typeset)}\\[2pt]%
  \rule{\linewidth}{0.4pt}%
  {\setlength{\parskip}{0pt}\setlength{\parindent}{0pt}\linespread{0.9}\fontsize{7pt}{8pt}\selectfont\input{examples/_generated/ex03_cnl.tex}}%
  \end{minipage}\hfill%
  \begin{minipage}[t]{0.5\textwidth}%
  \textbf{Hand-written Lean 4 proof}\\[2pt]%
  \rule{\linewidth}{0.4pt}%
  \lstinputlisting[language=Lean4]{examples/_generated/ex03_proof.lean}%
  \end{minipage}%
  \par\smallskip%
  \noindent{\small miniF2F \texttt{amc12a-2008-p8}. A four-step algebraic chain (\,$y=1$\,, $x^2=2$, $x=\sqrt{2}$, $x^3=2\sqrt{2}$). The CNL is four ``Then\ldots'' lines. The Lean proof factors $y^3-1$ to extract $y=1$, uses \texttt{Real.sq\_sqrt} to bridge $x^2=(\sqrt 2)^2$, and expands the cube as $(\sqrt 2)^2 \cdot \sqrt 2$.}\par\medskip%

  \par\addvspace{\medskipamount}%
  \noindent%
  \begin{minipage}[t]{0.47\textwidth}%
  \textbf{Visored CNL input (typeset)}\\[2pt]%
  \rule{\linewidth}{0.4pt}%
  {\setlength{\parskip}{0pt}\setlength{\parindent}{0pt}\linespread{0.9}\fontsize{7pt}{8pt}\selectfont\input{examples/_generated/ex05_cnl.tex}}%
  \end{minipage}\hfill%
  \begin{minipage}[t]{0.5\textwidth}%
  \textbf{Hand-written Lean 4 proof}\\[2pt]%
  \rule{\linewidth}{0.4pt}%
  \lstinputlisting[language=Lean4,basicstyle=\ttfamily\scriptsize]{examples/_generated/ex05_proof.lean}%
  \end{minipage}%
  \par\smallskip%
  \noindent{\small miniF2F \texttt{mathd-algebra-13}. Partial fractions: instantiate the universal hypothesis at $x=1$ and $x=2$, solve the resulting $2\times 2$ linear system. The Lean writer must specialize \texttt{h}, discharge the inequality side conditions (\texttt{norm\_num}), clear denominators with \texttt{field\_simp; ring\_nf}, and close with \texttt{linarith}.}\par\medskip%

\subsection{Number theory}
\label{sec:ex-numbertheory}

  \par\addvspace{\medskipamount}%
  \noindent%
  \begin{minipage}[t]{0.47\textwidth}%
  \textbf{Visored CNL input (typeset)}\\[2pt]%
  \rule{\linewidth}{0.4pt}%
  {\setlength{\parskip}{0pt}\setlength{\parindent}{0pt}\linespread{0.9}\fontsize{7pt}{8pt}\selectfont\input{examples/_generated/ex02_cnl.tex}}%
  \end{minipage}\hfill%
  \begin{minipage}[t]{0.5\textwidth}%
  \textbf{Hand-written Lean 4 proof}\\[2pt]%
  \rule{\linewidth}{0.4pt}%
  \lstinputlisting[language=Lean4]{examples/_generated/ex02_proof.lean}%
  \end{minipage}%
  \par\smallskip%
  \noindent{\small miniF2F \texttt{mathd-numbertheory-169}. One-line CNL discharging a numeric identity $\gcd(20!, 200000) = 40000$. Lean uses \texttt{native\_decide}; the writer must know \texttt{decide} times out at compile time.}\par\medskip%

  \par\addvspace{\medskipamount}%
  \noindent%
  \begin{minipage}[t]{0.47\textwidth}%
  \textbf{Visored CNL input (typeset)}\\[2pt]%
  \rule{\linewidth}{0.4pt}%
  {\setlength{\parskip}{0pt}\setlength{\parindent}{0pt}\linespread{0.9}\fontsize{7pt}{8pt}\selectfont\input{examples/_generated/ex06_cnl.tex}}%
  \end{minipage}\hfill%
  \begin{minipage}[t]{0.5\textwidth}%
  \textbf{Hand-written Lean 4 proof}\\[2pt]%
  \rule{\linewidth}{0.4pt}%
  \lstinputlisting[language=Lean4]{examples/_generated/ex06_proof.lean}%
  \end{minipage}%
  \par\smallskip%
  \noindent{\small miniF2F \texttt{mathd-numbertheory-188}. Small numeric gcd; \texttt{decide} suffices. The Lean writer also has to pick between \texttt{decide} and \texttt{native\_decide}, a choice that depends on how the numeric kernel evaluates the goal and is therefore an extra Lean-specific decision compared with the CNL version.}\par\medskip%

  \par\addvspace{\medskipamount}%
  \noindent%
  \begin{minipage}[t]{0.47\textwidth}%
  \textbf{Visored CNL input (typeset)}\\[2pt]%
  \rule{\linewidth}{0.4pt}%
  {\setlength{\parskip}{0pt}\setlength{\parindent}{0pt}\linespread{0.9}\fontsize{7pt}{8pt}\selectfont\input{examples/_generated/ex07_cnl.tex}}%
  \end{minipage}\hfill%
  \begin{minipage}[t]{0.5\textwidth}%
  \textbf{Hand-written Lean 4 proof}\\[2pt]%
  \rule{\linewidth}{0.4pt}%
  \lstinputlisting[language=Lean4,basicstyle=\ttfamily\scriptsize]{examples/_generated/ex07_proof.lean}%
  \end{minipage}%
  \par\smallskip%
  \noindent{\small miniF2F \texttt{mathd-numbertheory-780}. Modular arithmetic with case analysis. The CNL stays in $\mathbb{N}$, but Lean's truncated subtraction makes the literal $\mathbb{N}$ transliteration vacuously false (\texttt{x - 6\textasciicircum 2 = 0} whenever $x<36$), so the Lean writer must know to lift to $\mathbb{Z}$ --- itself part of the lemma-vocabulary cost. The proof uses \texttt{Int.dvd\_of\_emod\_eq\_zero}, \texttt{Int.modEq\_iff\_dvd}, and \texttt{interval\_cases}.}\par\medskip%

  \par\addvspace{\medskipamount}%
  \noindent%
  \begin{minipage}[t]{0.47\textwidth}%
  \textbf{Visored CNL input (typeset)}\\[2pt]%
  \rule{\linewidth}{0.4pt}%
  {\setlength{\parskip}{0pt}\setlength{\parindent}{0pt}\linespread{0.9}\fontsize{7pt}{8pt}\selectfont\input{examples/_generated/ex08_cnl.tex}}%
  \end{minipage}\hfill%
  \begin{minipage}[t]{0.5\textwidth}%
  \textbf{Hand-written Lean 4 proof}\\[2pt]%
  \rule{\linewidth}{0.4pt}%
  \lstinputlisting[language=Lean4,basicstyle=\ttfamily\tiny]{examples/_generated/ex08_proof.lean}%
  \end{minipage}%
  \par\smallskip%
  \noindent{\small miniF2F \texttt{imo-1984-p2}. Polynomial identity over $\mathbb{Z}$ plus $7^7$ divisibility plus square-root bound. Lean Mathlib chain: \texttt{Prime.pow\_dvd\_of\_dvd\_mul\_left}, \texttt{IsIntegrallyClosed.pow\_dvd\_pow\_iff}, \texttt{Nat.sqrt\_lt'}. Each lemma has its own positivity / coprimality side conditions.}\par\medskip%

\subsection{Inequalities}
\label{sec:ex-inequalities}

%
  \par\addvspace{\medskipamount}%
  \noindent%
  \begin{minipage}[t]{0.47\textwidth}%
  \textbf{Visored CNL input (typeset)}\\[2pt]%
  \rule{\linewidth}{0.4pt}%
  {\setlength{\parskip}{0pt}\setlength{\parindent}{0pt}\linespread{0.9}\fontsize{7pt}{8pt}\selectfont\input{examples/_generated/ex10_cnl.tex}}%
  \end{minipage}\hfill%
  \begin{minipage}[t]{0.5\textwidth}%
  \textbf{Hand-written Lean 4 proof}\\[2pt]%
  \rule{\linewidth}{0.4pt}%
  \lstinputlisting[language=Lean4]{examples/_generated/ex10_proof.lean}%
  \end{minipage}%
  \par\smallskip%
  \noindent{\small miniF2F \texttt{algebra-apb4leq8ta4pb4}. $(a+b)^4 \le 8(a^4+b^4)$. The Lean writer must supply the right \texttt{sq\_nonneg} hints (e.g., $\texttt{sq\_nonneg}\,(a-b)$, $\texttt{sq\_nonneg}\,(a^2-b^2)$) for \texttt{nlinarith} to close.}\par\medskip%

  \par\addvspace{\medskipamount}%
  \noindent%
  \begin{minipage}[t]{0.47\textwidth}%
  \textbf{Visored CNL input (typeset)}\\[2pt]%
  \rule{\linewidth}{0.4pt}%
  {\setlength{\parskip}{0pt}\setlength{\parindent}{0pt}\linespread{0.9}\fontsize{7pt}{8pt}\selectfont\input{examples/_generated/ex11_cnl.tex}}%
  \end{minipage}\hfill%
  \begin{minipage}[t]{0.5\textwidth}%
  \textbf{Hand-written Lean 4 proof}\\[2pt]%
  \rule{\linewidth}{0.4pt}%
  \lstinputlisting[language=Lean4,basicstyle=\ttfamily\scriptsize]{examples/_generated/ex11_proof.lean}%
  \end{minipage}%
  \par\smallskip%
  \noindent{\small miniF2F \texttt{algebra-amgm-faxinrrp2msqrt2geq2mxm1div2x}. AM-GM with $\sqrt 2$ and $1/(2x)$ on $x>0$. The Lean writer needs \texttt{Real.sq\_sqrt}, the hint $\texttt{sq\_nonneg}\,(x\sqrt 2-1)$, and the \texttt{field\_simp; ring} bridge to clear $1/(2x)$.}\par\medskip%

\subsection{Induction, sums, and products}
\label{sec:ex-induction}

%
  \par\addvspace{\medskipamount}%
  \noindent%
  \begin{minipage}[t]{0.47\textwidth}%
  \textbf{Visored CNL input (typeset)}\\[2pt]%
  \rule{\linewidth}{0.4pt}%
  {\setlength{\parskip}{0pt}\setlength{\parindent}{0pt}\linespread{0.9}\fontsize{7pt}{8pt}\selectfont\input{examples/_generated/ex12_cnl.tex}}%
  \end{minipage}\hfill%
  \begin{minipage}[t]{0.5\textwidth}%
  \textbf{Hand-written Lean 4 proof}\\[2pt]%
  \rule{\linewidth}{0.4pt}%
  \lstinputlisting[language=Lean4]{examples/_generated/ex12_proof.lean}%
  \end{minipage}%
  \par\smallskip%
  \noindent{\small miniF2F \texttt{induction\_sum\_1oktkp1}. Telescoping sum $\sum_{k=0}^{n-1}\frac{1}{(k+1)(k+2)} = \frac{n}{n+1}$. Lean induction on $n$, \texttt{Finset.sum\_range\_succ}, \texttt{field\_simp}, \texttt{ring}.}\par\medskip%

  \par\addvspace{\medskipamount}%
  \noindent%
  \begin{minipage}[t]{0.47\textwidth}%
  \textbf{Visored CNL input (typeset)}\\[2pt]%
  \rule{\linewidth}{0.4pt}%
  {\setlength{\parskip}{0pt}\setlength{\parindent}{0pt}\linespread{0.9}\fontsize{7pt}{8pt}\selectfont\input{examples/_generated/ex13_cnl.tex}}%
  \end{minipage}\hfill%
  \begin{minipage}[t]{0.5\textwidth}%
  \textbf{Hand-written Lean 4 proof}\\[2pt]%
  \rule{\linewidth}{0.4pt}%
  \lstinputlisting[language=Lean4,basicstyle=\ttfamily\scriptsize]{examples/_generated/ex13_proof.lean}%
  \end{minipage}%
  \par\smallskip%
  \noindent{\small miniF2F \texttt{amc12a-2008-p4}. Telescoping product $\prod_{k=1}^{501}\frac{4k+4}{4k} = 502$. Lean uses \texttt{Finset.prod\_Icc\_succ\_top} for the structural induction.}\par\medskip%

\subsection{Functional equations}
\label{sec:ex-functional}

  \par\addvspace{\medskipamount}%
  \noindent%
  \begin{minipage}[t]{0.47\textwidth}%
  \textbf{Visored CNL input (typeset)}\\[2pt]%
  \rule{\linewidth}{0.4pt}%
  {\setlength{\parskip}{0pt}\setlength{\parindent}{0pt}\linespread{0.9}\fontsize{7pt}{8pt}\selectfont\input{examples/_generated/ex14_cnl.tex}}%
  \end{minipage}\hfill%
  \begin{minipage}[t]{0.5\textwidth}%
  \textbf{Hand-written Lean 4 proof}\\[2pt]%
  \rule{\linewidth}{0.4pt}%
  \lstinputlisting[language=Lean4,basicstyle=\ttfamily\scriptsize]{examples/_generated/ex14_proof.lean}%
  \end{minipage}%
  \par\smallskip%
  \noindent{\small miniF2F \texttt{amc12a-2009-p9}. Polynomial coefficient identification: $f(x+3)=3x^2+7x+4$ and $f(x)=ax^2+bx+c$, find $a+b+c$. Lean specializes both hypotheses at $x=-2$ and $x=1$, closes with \texttt{linarith}.}\par\medskip%

  \par\addvspace{\medskipamount}%
  \noindent%
  \begin{minipage}[t]{0.47\textwidth}%
  \textbf{Visored CNL input (typeset)}\\[2pt]%
  \rule{\linewidth}{0.4pt}%
  {\setlength{\parskip}{0pt}\setlength{\parindent}{0pt}\linespread{0.9}\fontsize{7pt}{8pt}\selectfont\input{examples/_generated/ex15_cnl.tex}}%
  \end{minipage}\hfill%
  \begin{minipage}[t]{0.5\textwidth}%
  \textbf{Hand-written Lean 4 proof}\\[2pt]%
  \rule{\linewidth}{0.4pt}%
  \lstinputlisting[language=Lean4,basicstyle=\ttfamily\scriptsize]{examples/_generated/ex15_proof.lean}%
  \end{minipage}%
  \par\smallskip%
  \noindent{\small miniF2F \texttt{mathd-algebra-422}. $\sigma(\sigma^{-1}(x))=x$ trick. Lean's bijection layer requires \texttt{Function.Bijective.surjective} and \texttt{Function.LeftInverse.rightInverse\_of\_surjective} to extract the cancellation; the algebraic close is \texttt{linarith}.}\par\medskip%

\subsection{Sequences and recurrences}
\label{sec:ex-sequences}

  \par\addvspace{\medskipamount}%
  \noindent%
  \begin{minipage}[t]{0.47\textwidth}%
  \textbf{Visored CNL input (typeset)}\\[2pt]%
  \rule{\linewidth}{0.4pt}%
  {\setlength{\parskip}{0pt}\setlength{\parindent}{0pt}\linespread{0.9}\fontsize{7pt}{8pt}\selectfont\input{examples/_generated/ex17_cnl.tex}}%
  \end{minipage}\hfill%
  \begin{minipage}[t]{0.5\textwidth}%
  \textbf{Hand-written Lean 4 proof}\\[2pt]%
  \rule{\linewidth}{0.4pt}%
  \lstinputlisting[language=Lean4,basicstyle=\ttfamily\scriptsize]{examples/_generated/ex17_proof.lean}%
  \end{minipage}%
  \par\smallskip%
  \noindent{\small miniF2F \texttt{amc12a-2017-p7}. Recurrence with parity, induct on odd indices $f(2k+1) = 2k+2$. Lean uses \texttt{Nat.succ\_eq\_add\_one}, \texttt{push\_cast}, and \texttt{omega}.}\par\medskip%

\subsection{Trigonometry}
\label{sec:ex-trig}

  \par\addvspace{\medskipamount}%
  \noindent%
  \begin{minipage}[t]{0.47\textwidth}%
  \textbf{Visored CNL input (typeset)}\\[2pt]%
  \rule{\linewidth}{0.4pt}%
  {\setlength{\parskip}{0pt}\setlength{\parindent}{0pt}\linespread{0.9}\fontsize{7pt}{8pt}\selectfont\input{examples/_generated/ex19_cnl.tex}}%
  \end{minipage}\hfill%
  \begin{minipage}[t]{0.5\textwidth}%
  \textbf{Hand-written Lean 4 proof}\\[2pt]%
  \rule{\linewidth}{0.4pt}%
  \lstinputlisting[language=Lean4,basicstyle=\ttfamily\scriptsize]{examples/_generated/ex19_proof.lean}%
  \end{minipage}%
  \par\smallskip%
  \noindent{\small miniF2F \texttt{imo-1965-p1}. Inverse-trig reasoning. Lean Mathlib names: \texttt{Real.cos\_pi\_div\_four}, \texttt{Real.arccos\_cos}, \texttt{Real.arccos\_le\_arccos}. The interval $[\pi/4, 7\pi/4]$ has to be the canonical range for \texttt{arccos\_cos} to unfold.}\par\medskip%

\subsection{Logarithm and exponential}
\label{sec:ex-log}

  \par\addvspace{\medskipamount}%
  \noindent%
  \begin{minipage}[t]{0.47\textwidth}%
  \textbf{Visored CNL input (typeset)}\\[2pt]%
  \rule{\linewidth}{0.4pt}%
  {\setlength{\parskip}{0pt}\setlength{\parindent}{0pt}\linespread{0.9}\fontsize{7pt}{8pt}\selectfont\input{examples/_generated/ex20_cnl.tex}}%
  \end{minipage}\hfill%
  \begin{minipage}[t]{0.5\textwidth}%
  \textbf{Hand-written Lean 4 proof}\\[2pt]%
  \rule{\linewidth}{0.4pt}%
  \lstinputlisting[language=Lean4]{examples/_generated/ex20_proof.lean}%
  \end{minipage}%
  \par\smallskip%
  \noindent{\small miniF2F \texttt{mathd-algebra-245}. Exponent simplification chain. Lean steps: \texttt{inv\_div}, \texttt{pow\_succ}, \texttt{mul\_pow}, \texttt{pow\_mul}, \texttt{one\_div}, \texttt{inv\_inv}; one named lemma per CNL line.}\par\medskip%

  \par\addvspace{\medskipamount}%
  \noindent%
  \begin{minipage}[t]{0.47\textwidth}%
  \textbf{Visored CNL input (typeset)}\\[2pt]%
  \rule{\linewidth}{0.4pt}%
  {\setlength{\parskip}{0pt}\setlength{\parindent}{0pt}\linespread{0.9}\fontsize{7pt}{8pt}\selectfont\input{examples/_generated/ex21_cnl.tex}}%
  \end{minipage}\hfill%
  \begin{minipage}[t]{0.5\textwidth}%
  \textbf{Hand-written Lean 4 proof}\\[2pt]%
  \rule{\linewidth}{0.4pt}%
  \lstinputlisting[language=Lean4,basicstyle=\ttfamily\scriptsize]{examples/_generated/ex21_proof.lean}%
  \end{minipage}%
  \par\smallskip%
  \noindent{\small miniF2F \texttt{aime-1984-p5}. Log chain. The Lean version exercises a standard Mathlib path: \texttt{Real.log\_pow}, \texttt{Real.log\_mul} (with \texttt{ne\_of\_gt} side conditions for each factor), \texttt{Real.log\_pos} on $1<2$, and \texttt{Real.log\_injOn\_pos} to cancel the outer log. The CNL version is a chain of equalities; the Lean version makes every positivity hypothesis explicit.}\par\medskip%

\subsection{Set and membership}
\label{sec:ex-set}

  \par\addvspace{\medskipamount}%
  \noindent%
  \begin{minipage}[t]{0.47\textwidth}%
  \textbf{Visored CNL input (typeset)}\\[2pt]%
  \rule{\linewidth}{0.4pt}%
  {\setlength{\parskip}{0pt}\setlength{\parindent}{0pt}\linespread{0.9}\fontsize{7pt}{8pt}\selectfont\input{examples/_generated/ex24_cnl.tex}}%
  \end{minipage}\hfill%
  \begin{minipage}[t]{0.5\textwidth}%
  \textbf{Hand-written Lean 4 proof}\\[2pt]%
  \rule{\linewidth}{0.4pt}%
  \lstinputlisting[language=Lean4]{examples/_generated/ex24_proof.lean}%
  \end{minipage}%
  \par\smallskip%
  \noindent{\small miniF2F \texttt{mathd-numbertheory-221}. Set with divisor-count $=3$ (squares of primes) cardinality. The Lean writer rewrites the set as a \texttt{Finset.filter} over \texttt{Finset.range 1000} and closes with \texttt{native\_decide}; only works because the bound is a concrete numeral.}\par\medskip%

  \par\addvspace{\medskipamount}%
  \noindent%
  \begin{minipage}[t]{0.47\textwidth}%
  \textbf{Visored CNL input (typeset)}\\[2pt]%
  \rule{\linewidth}{0.4pt}%
  {\setlength{\parskip}{0pt}\setlength{\parindent}{0pt}\linespread{0.9}\fontsize{7pt}{8pt}\selectfont\input{examples/_generated/ex23_cnl.tex}}%
  \end{minipage}\hfill%
  \begin{minipage}[t]{0.5\textwidth}%
  \textbf{Hand-written Lean 4 proof}\\[2pt]%
  \rule{\linewidth}{0.4pt}%
  \lstinputlisting[language=Lean4,basicstyle=\ttfamily\scriptsize]{examples/_generated/ex23_proof.lean}%
  \end{minipage}%
  \par\smallskip%
  \noindent{\small miniF2F \texttt{mathd-algebra-224}. Set-builder with $\sqrt n \in (2, 7/2)$ rewritten as $n \in \{5,\ldots,12\}$. Lean uses \texttt{Real.lt\_sqrt}, \texttt{Real.sqrt\_lt'} to square out the radical, then \texttt{Set.ext} / \texttt{Finset.ext} / \texttt{Nat.card\_Icc} for the final cardinality.}\par\medskip%

%% file: examples/_generated/ex03_cnl.tex
\begin{example}
Let $x\in\mathbb{R}$.

Let $y\in\mathbb{R}$.

Assume $x>0$.

Assume $y>0$.

Assume $y^3=1$.

Assume $6x^2=2(6y^2)$.

The goal is to prove $x^3=2\sqrt{2}$.

Then $y=1$.

Then $x^2=2$.

Then $x=\sqrt{2}$.

Then $x^3=2\sqrt{2}$.
\end{example}

%% file: examples/_generated/ex05_cnl.tex
\begin{example}
Let $a\in\mathbb{R}$.

Let $b\in\mathbb{R}$.

Assume $\forall x\in\mathbb{R},\,(x\neq3 \land x\neq5) \rightarrow \frac{4x}{x^2-8x+15}=\frac{a}{x-3}+\frac{b}{x-5}$.

The goal is to prove $a=-6 \land b=10$.

We have $(1\neq3 \land 1\neq5) \rightarrow \frac{4}{1-8+15}=\frac{a}{1-3}+\frac{b}{1-5}$.

We have $1\neq3$.

We have $1\neq5$.

Then $\frac{4}{8}=\frac{a}{-2}+\frac{b}{-4}$.

Then $2a+b=-2$.

We have $(2\neq3 \land 2\neq5) \rightarrow \frac{8}{4-16+15}=\frac{a}{2-3}+\frac{b}{2-5}$.

We have $2\neq3$.

We have $2\neq5$.

Then $\frac{8}{3}=\frac{a}{-1}+\frac{b}{-3}$.

Then $3a+b=-8$.

Then $a=(3a+b)-(2a+b)=-8-(-2)=-6$.

Then $b=-2-2a=-2-2 \cdot (-6)=10$.

Then $a=-6 \land b=10$.
\end{example}

%% file: examples/_generated/ex02_cnl.tex
\begin{example}
The goal is to prove $\gcd(20!,200000)=40000$.

We have $\gcd(20!,200000)=40000$.
\end{example}

%% file: examples/_generated/ex06_cnl.tex
\begin{example}
The goal is to prove $\gcd(180,168)=12$.

We have $\gcd(180,168)=12$.
\end{example}

%% file: examples/_generated/ex07_cnl.tex
\begin{example}
Let $m\in\mathbb{N}$.

Let $x\in\mathbb{N}$.

Assume $10\le m\le 99$.

Assume $(6x)\bmod m=1$.

Assume $(x-6^2)\bmod m=0$.

The goal is to prove $m=43$.

We have $x\equiv 36 \pmod{m}$.

We have $6x\equiv 6\cdot 36 \pmod{m}$.

We have $6x\equiv 1 \pmod{m}$.

We have $6\cdot 36\equiv 1 \pmod{m}$.

We have $216\equiv 1 \pmod{m}$.

We have $m\mid 215$.

We prove $m=43$ by working through different cases :
\begin{itemize}
    \item Case $m=1$. Contradiction.

\item Case $m=5$. Contradiction.

\item Case $m=43$.

We have $m=43$.

\item Case $m=215$. Contradiction.

\end{itemize}
\end{example}

%% file: examples/_generated/ex08_cnl.tex
\begin{example}
Let $a\in\mathbb{N}$.

Let $b\in\mathbb{N}$.

Assume $0<a$.

Assume $0<b$.

Assume $7 \nmid a$.

Assume $7 \nmid b$.

Assume $7 \nmid (a+b)$.

Assume $7^7\mid ({(a+b)}^7-a^7-b^7)$.

The goal is to prove $a+b\ge 19$.

We have ${(a+b)}^7 - a^7 - b^7 = 7ab(a+b){(a^2+ab+b^2)}^2$.

We have $7^7 \mid 7ab(a+b){(a^2+ab+b^2)}^2$.

We have $7^6 \mid ab(a+b){(a^2+ab+b^2)}^2$.

We have $7 \nmid ab$.

We have $7 \nmid ab(a+b)$.

We have $7^6 \mid {(a^2+ab+b^2)}^2$.

We have $7^3 \mid a^2+ab+b^2$.

We have $a \geq 1$.

We have $b \geq 1$.

We have $a^2 \geq 1$.

We have $a^2+ab+b^2 \geq 1$.

We have $a^2+ab+b^2 \geq 343$.

We have $ab \geq 0$.

We have ${(a+b)}^2 = a^2 + 2ab + b^2$.

We have ${(a+b)}^2 \geq a^2 + ab + b^2$.

We have ${(a+b)}^2 \geq 343$.

We have $a+b > 0$.

We have $a+b \geq 19$.
\end{example}

%% file: examples/_generated/ex10_cnl.tex
\begin{example}
Let $a,b\in\mathbb{R}$.

Assume $a>0$.

Assume $b>0$.

The goal is to prove ${(a+b)}^4\le8(a^4+b^4)$.

We have ${(a+b)}^4 = a^4 + 4a^3b + 6a^2b^2 + 4ab^3 + b^4$.

We have $8(a^4+b^4) - {(a+b)}^4 = 7a^4 - 4a^3b - 6a^2b^2 - 4ab^3 + 7b^4$.

We have $7a^4 - 4a^3b - 6a^2b^2 - 4ab^3 + 7b^4 = {(a-b)}^2(7a^2 + 10ab + 7b^2)$.

We have ${(a-b)}^2 \ge 0$.

We have $a^2 > 0$.

We have $b^2 > 0$.

We have $ab > 0$.

We have $7a^2 > 0$.

We have $7b^2 > 0$.

We have $10ab > 0$.

We have $7a^2 + 10ab + 7b^2 > 0$.

We have ${(a-b)}^2(7a^2 + 10ab + 7b^2) \ge 0$.

We have $8(a^4+b^4) - {(a+b)}^4 \ge 0$.

Then ${(a+b)}^4\le8(a^4+b^4)$.
\end{example}

%% file: examples/_generated/ex11_cnl.tex
\begin{example}
Let $x\in\mathbb{R}$.

Assume $x>0$.

The goal is to prove $2-\sqrt{2}\ge2-x-\frac{1}{2x}$.

We have $x + \frac{1}{2x} - \sqrt{2} = {\left(\sqrt{x} - \frac{1}{\sqrt{2x}}\right)}^2$.

We have ${\left(\sqrt{x} - \frac{1}{\sqrt{2x}}\right)}^2 \ge 0$.

We have $x + \frac{1}{2x} - \sqrt{2} \ge 0$.

We have $x + \frac{1}{2x} \ge \sqrt{2}$.

We have $2 - x - \frac{1}{2x} \le 2 - \sqrt{2}$.

Then $2-\sqrt{2}\ge2-x-\frac{1}{2x}$.
\end{example}

%% file: examples/_generated/ex12_cnl.tex
\begin{example}
Let $n\in\mathbb{N}$.

The goal is to prove $\sum_{k=0}^{n-1}\frac{1}{(k+1)(k+2)}=\frac{n}{n+1}$.

We have $\forall k \in \mathbb{N},\, \frac{1}{(k+1)(k+2)} = \frac{1}{k+1} - \frac{1}{k+2}$.

We have $\sum_{k=0}^{n-1}\frac{1}{(k+1)(k+2)} = \sum_{k=0}^{n-1}\left(\frac{1}{k+1} - \frac{1}{k+2}\right)$.

We have $\sum_{k=0}^{n-1}\left(\frac{1}{k+1} - \frac{1}{k+2}\right) = 1 - \frac{1}{n+1}$.

Then $\sum_{k=0}^{n-1}\frac{1}{(k+1)(k+2)} = 1 - \frac{1}{n+1}$.

We have $1 - \frac{1}{n+1} = \frac{n}{n+1}$.

Then $\sum_{k=0}^{n-1}\frac{1}{(k+1)(k+2)}=\frac{n}{n+1}$.
\end{example}

%% file: examples/_generated/ex13_cnl.tex
\begin{example}
The goal is to prove $\prod_{k=1}^{501} \frac{4k+4}{4k}=502$.

We have $\forall k \in \mathbb{N},\, k > 0 \rightarrow \frac{4k+4}{4k} = \frac{k+1}{k}$.

We have $\prod_{k=1}^{501} \frac{4k+4}{4k} = \prod_{k=1}^{501} \frac{k+1}{k}$.

We have $\prod_{k=1}^{501} \frac{k+1}{k} = \frac{502}{1}$.

We have $\frac{502}{1} = 502$.

Then $\prod_{k=1}^{501} \frac{4k+4}{4k}=502$.
\end{example}

%% file: examples/_generated/ex14_cnl.tex
\begin{example}
Let $a\in\mathbb{R}$.

Let $b\in\mathbb{R}$.

Let $c\in\mathbb{R}$.

Let $f:\mathbb{R}\to\mathbb{R}$ be a function.

Assume $\forall x \in \mathbb{R},\,f(x+3)=3x^2+7x+4$.

Assume $\forall x \in \mathbb{R},\,f(x)=ax^2+bx+c$.

The goal is to prove $a+b+c=2$.

We have $\forall x \in \mathbb{R},\,f(x+3) = a{(x+3)}^2 + b(x+3) + c$.

We have $\forall x \in \mathbb{R},\,3x^2+7x+4 = a{(x+3)}^2 + b(x+3) + c$.

We have $\forall x \in \mathbb{R},\,a{(x+3)}^2 + b(x+3) + c = ax^2 + 6ax + 9a + bx + 3b + c$.

We have $\forall x \in \mathbb{R},\,3x^2+7x+4 = ax^2 + 6ax + 9a + bx + 3b + c$.

We have $\forall x \in \mathbb{R},\,ax^2 + 6ax + 9a + bx + 3b + c = ax^2 + (6a+b)x + (9a+3b+c)$.

We have $\forall x \in \mathbb{R},\,3x^2+7x+4 = ax^2 + (6a+b)x + (9a+3b+c)$.

We have $a = 3$.

We have $6a + b = 7$.

We have $b = -11$.

We have $9a + 3b + c = 4$.

We have $c = 10$.

We have $a + b + c = 2$.
\end{example}

%% file: examples/_generated/ex15_cnl.tex
\begin{example}
Let $x\in\mathbb{R}$.

Let $\sigma: \mathbb{R} \rightarrow \mathbb{R}$ be a function.

Assume $\sigma$ is a bijection.

Assume $\forall t\in\mathbb{R},\, \sigma(t)=5t-12$.

Assume $\sigma(x+1)=\sigma^{-1}(x)$.

The goal is to prove $x=\frac{47}{24}$.

We have $\sigma(x+1) = 5(x+1) - 12$.

Then $\sigma(x+1) = 5x - 7$.

Then $\sigma^{-1}(x) = 5x - 7$.

We have $\sigma(5x-7) = 5(5x-7) - 12$.

Then $\sigma(5x-7) = 25x - 47$.

We have $\sigma(\sigma^{-1}(x)) = x$.

Then $25x - 47 = x$.

Then $24x = 47$.

Then $x = 47/24$.
\end{example}

%% file: examples/_generated/ex17_cnl.tex
\begin{example}
Let $f:\mathbb{N}\to\mathbb{R}$ be a function.

Assume $f(1)=2$.

Assume $\forall n\in\mathbb{N},\, n>1 \land n\bmod2=0 \implies f(n)=f(n-1)+1$.

Assume $\forall n\in\mathbb{N},\, n>1 \land n\bmod2=1 \implies f(n)=f(n-2)+2$.

The goal is to prove $f(2017)=2018$.

Let $k\in\mathbb{N}$.

We prove $f(2k+1) = 2k + 2$ by induction on $k$ :

\begin{itemize}

\item Case $k=0$.

We have $f(1) = 2$.

We have $2 \cdot 0 + 2 = 2$.

Then $f(2 \cdot 0+1) = 2 \cdot 0 + 2$.

\item Case $k \ge 0$.

Assume $f(2k+1) = 2k + 2$.

The goal is to prove $f(2(k+1)+1) = 2(k+1) + 2$.

We have $2(k+1)+1 = 2k+3$.

We have $2k+3 > 1$.

We have $2k+3 \equiv 1 \pmod{2}$.

Then $(2k+3) \bmod 2 = 1$.

We have $f(2k+3) = f((2k+3)-2) + 2$.

We have $(2k+3) - 2 = 2k+1$.

Then $f(2k+3) = f(2k+1) + 2$.

We have $f(2k+1) = 2k + 2$.

Then $f(2k+3) = (2k+2) + 2$.

We have $(2k+2) + 2 = 2k + 4$.

We have $2k + 4 = 2(k+1) + 2$.

Then $f(2(k+1)+1) = 2(k+1) + 2$.

\end{itemize}

We have $2017 = 2 \cdot 1008 + 1$.

We have $f(2 \cdot 1008+1) = 2 \cdot 1008 + 2$.

We have $2 \cdot 1008 + 2 = 2018$.

Then $f(2017) = 2018$.
\end{example}

%% file: examples/_generated/ex19_cnl.tex
\begin{example}
Let $x\in\mathbb{R}$.

Assume $0\le x\le 2\pi$.

Assume $2\cos x\le\Bigl|\sqrt{1+\sin(2x)}-\sqrt{1-\sin(2x)}\Bigr|\le\sqrt{2}$.

The goal is to prove $\frac{\pi}{4}\le x\le\frac{7\pi}{4}$.

We have $2\cos x \le \sqrt{2}$.

We have $\cos x \le \frac{\sqrt{2}}{2}$.

We have $\cos\left(\frac{\pi}{4}\right) = \frac{\sqrt{2}}{2}$.

We have $\cos\left(\frac{7\pi}{4}\right) = \frac{\sqrt{2}}{2}$.

Then $\frac{\pi}{4} \le x \le \frac{7\pi}{4}$.
\end{example}

%% file: examples/_generated/ex20_cnl.tex
\begin{example}
Let $x\in\mathbb{R}$.

Assume $x\neq0$.

The goal is to prove ${\left(\frac{4}{x}\right)}^{-1}{\left(\frac{3x^3}{x}\right)}^{2}{\left(\frac{1}{2x}\right)}^{-3}=18x^8$.

We have ${\left(\frac{4}{x}\right)}^{-1}{\left(\frac{3x^3}{x}\right)}^{2}{\left(\frac{1}{2x}\right)}^{-3}=18x^8$.
\end{example}

%% file: examples/_generated/ex21_cnl.tex
\begin{example}
Let $a\in\mathbb{R}$.

Let $b\in\mathbb{R}$.

Assume $a>0$.

Assume $b>0$.

Assume $\log_8 a+\log_4(b^2)=5$.

Assume $\log_8 b+\log_4(a^2)=7$.

The goal is to prove $ab=512$.

We have $\frac{1}{3}\log_2 a + \frac{1}{2} \log_2{(b^2)} = \frac{1}{3}\log_2 a + \frac{2}{2} \log_2 b = \frac{1}{3}\log_2 a + \log_2 b = 5$.

We have $\frac{1}{3}\log_2 b + \frac{1}{2}\log_2{(a^2)} = \frac{1}{3}\log_2 b + \frac{2}{2}\log_2 a = \frac{1}{3}\log_2 b + \log_2 a = 7$.

We have $3(\frac{1}{3}\log_2 a + \log_2 b) = \log_2 a + 3\log_2 b = \log_2 a + \log_2{(b^3)} = \log_2(ab^3) = 15$.

We have $ab^3 = 2^{15}$.

We have $3(\frac{1}{3}\log_2 b + \log_2 a) = \log_2 b + 3\log_2 a = \log_2 b + \log_2{(a^3)} = \log_2(ba^3) = 21$.

We have $ba^3 = 2^{21}$.

We have $(ab^3)(ba^3) = {(ab)}^4 = 2^{15} 2^{21} = 2^{36}$.

We have $ab = 2^{36/4} = 2^9 = 512$.
\end{example}

%% file: examples/_generated/ex24_cnl.tex
\begin{example}
Let $S=\{x\in\mathbb{N}\mid 0<x<1000 \land \lvert \{ d \in \mathbb{N} \mid d \mid x \} \rvert=3\}$.

The goal is to prove $\lvert S \rvert=11$.

We have $S = \{4, 9, 25, 49, 121, 169, 289, 361, 529, 841, 961\}$.

We have $\lvert S \rvert = 11$.
\end{example}

%% file: examples/_generated/ex23_cnl.tex
\begin{example}
Let $S=\{n\in\mathbb{N}\mid\sqrt{n}<\frac{7}{2} \land \sqrt{n}>2\}$.

The goal is to prove $|S|=8$.

Let $n \in \mathbb{N}$.

Then $\sqrt{n} > 2 \iff n > 4$.

Then $\sqrt{n} < \frac{7}{2} \iff n < \frac{49}{4}$.

Then $n > 4 \iff n \ge 5$.

We have $\frac{49}{4} = 12 + \frac{1}{4}$.

Then $n < \frac{49}{4} \iff n \le 12$.

Then $\sqrt{n}<\frac{7}{2} \land \sqrt{n}>2 \iff 5 \le n \le 12$.

Then $S = \{n \in \mathbb{N} \mid 5 \le n \le 12\}$.

We have $|\{n \in \mathbb{N} \mid 5 \le n \le 12\}| = 8$.

Then $|S| = 8$.
\end{example}